% Comment to ward off arxiv errors
%auto-ignore 

\input eplain
% Comment to ward off arxiv errors
%auto-ignore 

\def\CurrentFontFile{}
\newdimen\DefaultFontSize
\DefaultFontSize=10pt
\newdimen\CurrentFontSize
\def\SetFont#1{\SetFontSpec{#1}{\CurrentFontSize}}

\def\SetFontSpec#1#2{\edef\CurrentFontFile{#1}\CurrentFontSize=#2\font\ActiveFont#1 at\CurrentFontSize\ActiveFont}
\SetFontSpec{ptmr}{\DefaultFontSize}

\def\rm{\fam0\SetFont{\BaseFont}}
\def\it{\fam\itfam\SetFont{\ItalicFont}}
\def\sl{\fam\slfam\SetFont{\SlantedFont}}
\def\bf{\fam\bffam\SetFont{\BoldFont}}

\def\SetTextFontGroup#1#2{\def\InstallProc{\csname InstallFontGroup#1\endcsname}\expandafter
\ifx\InstallProc\relax\errmessage{Unable to install font group #1}\else
\InstallProc{#2}\fi}

\def\SetMathFontGroup#1#2{\def\InstallProc{\csname InstallMathFontGroup#1\endcsname}\expandafter
\ifx\InstallProc\relax\errmessage{Unable to install math font group #1}\else
\InstallProc{#2}\fi}

\def\InstallMathFontGroupTimesOld#1{%
%\font\tf=ztmcmr at 12pt
%\font\sf=ztmcmr at 9pt
%\font\ssf=ztmcmr at 7pt
\font\tf=cmr12
\font\sf=cmr9
\font\ssf=cmr7
\textfont0=\tf 
\scriptfont0=\sf
\scriptscriptfont0=\ssf
%\font\tf=ztmcmrm at 12pt
%\font\sf=ztmcmrm at 9pt
%\font\ssf=ztmcmrm at 7pt
\font\tf=cmmi12
\font\sf=cmmi9
\font\ssf=cmmi7
\skewchar\tf='177
\skewchar\sf='177
\skewchar\ssf='177
\textfont1=\tf 
\scriptfont1=\sf
\scriptscriptfont1=\ssf
\font\tfs=cmsy10 at 12pt
\font\sfs=cmsy9
\font\ssfs=cmsy7
\skewchar\tfs='60
\skewchar\sfs='60
\skewchar\ssfs='60
\textfont2=\tfs 
\scriptfont2=\sfs
\scriptscriptfont2=\ssfs
\textfont3=\tenex
\scriptfont3=\tenex
\scriptscriptfont3=\tenex
\font\tfsa=txsya at 12pt
\font\sfsa=txsya at 9pt
\font\ssfsa=txsya at 7pt
\textfont\msafam=\tfsa
\scriptfont\msafam=\sfsa
\scriptscriptfont\msafam=\ssfsa
\font\tfsb=txsyb at 12pt
\font\sfsb=txsyb at 9pt
\font\ssfsb=txsyb at 7pt
\textfont\msbfam=\tfsb
\scriptfont\msbfam=\sfsb
\scriptscriptfont\msbfam=\ssfsb}

\newdimen\mtsize
\newdimen\mssize
\newdimen\msssize
\newfam\msafam
\newfam\msbfam
\newfam\rsfsfam
\newfam\zcfam
\newfam\eufam

\def\Bbb#1{{\fam\msbfam\relax#1}}

\SetTextFontGroup{Times}{\DefaultFontSize}
\SetMathFontGroup{Times}{\DefaultFontSize}

\input amssym.tex
\DefaultFontSize=12pt
\SetTextFontGroup{Times}{12pt}
\SetMathFontGroup{Times}{12pt}

\def\InstallTimesMath#1{\SetMathFontGroup{Times}{#1}}

\vsize=8.6in
\hsize=6in
\hoffset 0.25 in
\voffset 0.25 in
\parskip = 8 pt
\parindent = 0pt
%\advance \baselineskip by 8pt

% High level formatting
\newcount\sectioncount
\newcount\subsectioncount

\def\seclabel#1{\definexref{#1}{\number\sectioncount}{sec}}

\def\bl{}
\def\el{}
\def\li#1{\par\penalty500\hangindent=4em
\hangafter=0\indent\llap{\SetFontSpec{ptmr}{12pt}#1\hskip 0.5em}}
\def\define#1{#1}%

\newif\ifdraft
\drafttrue
\def\Version#1{\draftfalse\gdef\versiondate{#1}}

\newif\iflatexing\latexingfalse
\newif\ifjournal\journalfalse

\def\section#1{\advance\sectioncount by 1 \subsectioncount = 0 \proccount=0
\vskip 10pt plus 6pt minus 2pt \goodbreak
{\InstallTimesMath{16pt}\SetFontSpec{pplb}{16pt}\number\sectioncount. #1\par}}

\def\subsection#1{\advance\subsectioncount by 1
\vskip 5pt plus 3pt \goodbreak
{\InstallTimesMath{13pt}\SetFontSpec{pplb}{13pt}
\number\sectioncount.\number\subsectioncount\ #1\par}}

\def\startmatter{\bgroup\noindent\vskip 2cm plus 0.5cm 
\baselineskip=12pt
\ifx\titletext{}\else{\raggedright\SetFontSpec{pplb}{20pt}\hbox to \hsize{\vbox{\titletext}}}\bigskip\fi
\SetFontSpec{ptmr}{12pt}%
\InstallTimesMath{12pt}%
\ifx\authortext{}\else{\bf \authortext}\hfill
\vskip 0pt{\authorhome\hfill\parskip=0pt\vskip 0pt\email\authoremail}
\smallskip\fi
\ifdraft Draft: \today\else\versiondate\fi\vskip 1cm
\ifx\abstracttext{}\else{{\SetFontSpec{pplb}{10pt}Abstract. 
}\SetFontSpec{ptmr}{10pt}\abstracttext\smallskip}\fi
\egroup\advance \baselineskip by 4pt}
\let\endmatter\bye

\def\title#1{\global\def\titletext{#1}}
\def\titletext{}
\def\Author#1#2#3{\global\def\authortext{#1}\gdef\authorhome{#2}\gdef\authoremail{#3}}
\def\authortext{}
\def\authorhome{}
\def\Abstract#1{\global\def\abstracttext{#1}}
\def\abstracttext{}
\def\headtitle{{\SetFontSpec{ptmri}{10pt}\titletext\ifdraft\ (Draft: \today)\fi}} 
\def\headauthor{{\SetFontSpec{ptmri}{10pt}\authortext}} 
\def\email#1{{E-mail: #1}}%

\headline={\ifnum\pageno>1\SetFontSpec{ptmr}{10pt}%
\ifx\titletext{}\folio.\hfill\else
\ifodd\pageno\headtitle\hfill\folio\else
\folio\hfill\headauthor\fi\fi\else\hfil\fi}
\nopagenumbers

\catcode`@=11 % at signs are  letters
\def\makeheadline{\vbox to\z@{\vskip-35\p@
  \line{\vbox to8.5\p@{}\the\headline}\vss}\nointerlineskip}
\catcode`@=12 % at signs are no longer letters

% ams bib stuff
\catcode`@=11 % at signs are  letters
\def\providecommand#1{
   \def\@commandname{#1}%
   \@getoptionalarg\@postprovidecommand
}
\def\@postprovidecommand#1{}
\catcode`@=12 % at signs are no longer letters
\def\textbf#1{{\bf#1}}

\newcount\proccount
\def\myproclaim#1{\advance\proccount by 1\medbreak\noindent{\bf #1.%
\enspace}\bgroup\sl}
\def\procnumber{\number\sectioncount.\number\proccount}

\def\endmyproclaim{\egroup\par
\ifdim\lastskip<\medskipamount \removelastskip\penalty55\medskip\fi}

\def\Proof{\endmyproclaim{\it Proof:\hskip 0.5em}}
\def\NoProof{\endmyproclaim}
\def\EOProof{\unskip\quad$\square$\goodbreak\par}

\def\propword{Proposition}
\def\proplabel#1{\definexref{#1}{\procnumber}{prop}}
\def\prop{\myproclaim{\propword\ \procnumber}\def\label{\proplabel}}

\def\theoremword{Theorem}
\def\theoremlabel#1{\definexref{#1}{\procnumber}{theorem}}
\def\thm{\myproclaim{\theoremword\ \procnumber}\def\label{\theoremlabel}}

\def\lemmaword{Lemma}
\def\lemmalabel#1{\definexref{#1}{\procnumber}{lemma}}
\def\lem{\myproclaim{\lemmaword\ \procnumber}\def\label{\lemmalabel}}

\def\corword{Corollary}
\def\corlabel#1{\definexref{#1}{\procnumber}{cor}}
\def\cor{\myproclaim{\corword\ \procnumber}\def\label{\corlabel}}

\def\Definition{\advance\proccount by 1\medbreak\noindent{\bf 
Definition \procnumber}\enspace\bgroup\sl}
\def\EODefinition{\egroup\par\bigskip}

\def\Remark{{\bf Remark.}\enspace}
\def\EORemark{\medskip\par}

% multiline math
\def\({$$\eqalign\begingroup}
\def\){\endgroup$$}

\def\Acknowledge{\bigskip{\bf Acknowledgements.}\enspace}
\def\EOAcknowledge{\par}

\catcode`@=11 % at signs are  letters

\def\eqaligntd#1{\null\,\vcenter\bgroup\openup\jot\m@th
  \ialign\bgroup\strut\hfil$\displaystyle{##}$&$\displaystyle{{}##}$&$\displaystyle{{}##}$\hfil
\crcr}
\def\endeqaligntd{\crcr\egroup\egroup\,}

\def\eqalignd{\null\,\vcenter\bgroup\openup\jot\m@th
  \ialign\bgroup\strut\hfil$\displaystyle{##}$&$\displaystyle{{}##}$\hfil
      \crcr}      
\def\endeqalignd{\crcr\egroup\egroup\,}

\def\eqalignnod{\displ@y \tabskip\centering
  \halign to\displaywidth\bgroup\hfil$\@lign\displaystyle{##}$\tabskip\z@skip
    &$\@lign\displaystyle{{}##}$\hfil\tabskip\centering
    &\llap{$\@lign##$}\tabskip\z@skip\crcr}
\def\endeqalignnod{\crcr\egroup}

\catcode`@=12 % at signs are no longer letters

\def\be{$$}
\def\ee{$$}
\def\bel#1{\gdef\eqlabel{#1}$$}
\def\eel{\eqdef{\eqlabel}$$}
\def\bea{\bgroup\let\\\cr$$\eqalignd}
\def\eea{\endeqalignd$$\egroup}
\def\beal#1{\bgroup\let\\\cr\gdef\eqlabel{#1}$$\eqalignd}
\def\eeal{\endeqalignd\eqdef{\eqlabel}$$\egroup}
\def\beat{\bgroup\let\\\cr$$\eqaligntd}
\def\eeat{\endeqaligntd$$\egroup}
\def\beatl#1{\bgroup\let\\\cr\gdef\eqlabel{#1}$$\eqaligntd}
\def\eeatl{\endeqaligntd\eqdef{\eqlabel}$$\egroup}
\def\bean{\bgroup\let\\\crcr\let\label\ealabel
$$\eqalignnod}
\def\eean{\endeqalignnod$$\egroup}
\def\ealabel#1{&\eqdef{#1}\crcr}

\let\ref\refn
\def\nonumber{}

% Some style types
%\def\emph{\SetFontSpec{phvb}{10pt}}
%\def\Emph{\SetFontSpec{phvb}{14pt}}

\def\sf{\SetFontSpec{phvr}{10pt}}

% Extra definitions

\def\til{\lower 0.75ex\hbox{\char126}}

% Sobolov spaces
\def\W#1#2{W^{{#1},{#2}}}

% Holder spaces

% Other math defs
\def\Lap{\Delta\,}

\def\emph#1{{\it#1}}

\def\references{\section{References}}

\input epsf

\newcount\figurecount
\def\figlabel#1{\definexref{#1}{\number\figurecount}{fig}}
\def\fig#1#2{\advance\figurecount by 1\noindent\vskip\parskip\bigskip
{\def\label{\figlabel}\vbox{\hbox to \hsize{\hfil#1\hfil}\par\hfil
Figure \number\figurecount: #2\hfil\bigskip}}}

% Comment to ward off arxiv errors
%auto-ignore 
\def\bnorm#1{\left|\left|#1\right|\right|}

\def\gflat{{\overline g} }
\def\Reals{{\Bbb R}}

\def\Rn{\Reals^n}
\def\abs#1{\left|#1\right|}

\def\loc{_{\rm loc}}

\def\cpct{_{\rm c}}

\def\gchi{\raise 0.4 ex\hbox{$\chi$}}

\def\calL{{\cal L}}

\def\calP{{\cal P}}

\def\calS{{\cal S}}
\def\calF{{\cal F}}
\def\calG{{\cal G}}
\def\diff{{\rm d}\,}

\def\del{_\delta}

\def\srho{_{\rho}}

\def\tr{\mathop{\rm tr}}
\def\div{\mathop{\rm div}}
\def\supp{\mathop{\rm supp}}

\def\ck{{\Bbb L}}
\def\vl{\Delta_{\Bbb L}}

\def\Bflat{\overline{B}}
\def\vlflat{\overline{\Delta}_{\Bbb L}}
\def\ckflat{\overline{\Bbb L}}

\def\dV{\;dV}

\def\dimc{\kappa}
\def\expfrac#1#2{{#1\over#2}}
\def\ip<#1,#2>{\left<#1,#2\right>}

\def\low{_{\rm low}}
\def\H#1{H^{#1}}

\def\cpctsubset{\subset\subset}
\def\Htt{{H^{s-1}_{\delta-1}}}
\def\Hmt{{H^{s}_{\delta}}}

\def\ra{\rightarrow}

\def\divk{\mathop{ {\rm div}_{k}}}
\def\im{\mathop{{\rm im}}}

\Version{May 17, 2004}
\draftfalse
\title{\raggedright Rough Solutions of the Einstein Constraint Equations}
\ifjournal
\Author{David Maxwell}{Department of Mathematics, University of Washington,  
Seattle, Washington, 98195-4350, USA}{dmaxwell@math.washington.edu}
\else
\Author{David Maxwell}{University of Washington}{dmaxwell@math.washington.edu}
\fi
\Abstract{We construct low regularity solutions of the vacuum
Einstein constraint equations.  In particular, on 3-manifolds we obtain
solutions with metrics in $H^s\loc$ with
$s>{3\over 2}$.  The theory of maximal asymptotically Euclidean
solutions of the constraint equations descends completely the low
regularity setting.  Moreover, every rough, maximal, asymptotically
Euclidean solution can be approximated in an appropriate topology by
smooth solutions.  These results have application in an existence
theorem for rough solutions of the Einstein evolution equations.  }
\startmatter

\section{Introduction}
The Cauchy problem in general relativity can be formulated as a second-order
system of hyperbolic PDEs known as the Einstein equations.
Initial data for this problem is a 3-dimensional Riemannian manifold
$(M,g)$, and a symmetric $(0,2)$-tensor $K$ which plays the role of the
time derivative of $g$.  More precisely, $K$ specifies the second
fundamental form of an isometric embedding of $(M,g)$ into an ambient
4-dimensional Lorentzian manifold to be determined by solving the
Cauchy problem.  For the vacuum Cauchy problem, the ambient Lorentzian
manifold is Ricci-flat, and this imposes a compatibility condition on
the initial data known as the Einstein constraint equations.  These
are
\beal{constraints}
R-\abs{K}^2+\tr K ^2 &= 0\\
\div K -d\,\tr K &= 0,
\eeal
where the scalar curvature $R$ and all other quantities involving a
metric in \eqref{constraints} are computed with respect to $g$.

For any evolution problem, a natural question is to determine function
spaces in which it is well-posed.  The first local well-posedness
results for the Einstein equations were established in
\cite{cb-cauchy} for initial data $(g,K)\in C^5\times C^4$.  Subsequent
improvements lead to a local well-posedness result 
for the Einstein equations \cite{well-posed-hyp}
that requires initial data with 
$(g, K)\in H^{s}\loc\times H^{s-1}\loc$ with $s>5/2$ (we
ignore, for the moment, precise asymptotic conditions at infinity).
We will call a
solution with this last level of regularity a classical solution.
Recent work in the theory of nonlinear hyperbolic PDEs has lowered
the amount of regularity required.  Smith and Tataru
\cite{smith-tataru-lwp} have obtained local well-posedness for
certain nonlinear wave equations with initial data in $H^s\times H^{s-1}$
with $s>2$.  In the case of the vacuum Einstein equations, Klainerman
and Rodnianski \cite{kr-rough-annals} established an a-priori estimate
for the time of existence of a classical solution of the Einstein
equations in terms of the norm of $(\nabla g,K)$ in $H^{s-1}\times H^{s-1}$, 
again with $s>2$. These results should have lead to an existence theorem for
rough solutions, but the corresponding low regularity theory of the
constraint equations was not sufficiently well developed.  It was not
known if there existed any rough solutions of the constraint
equations.  Moreover, to pass from the a-priori estimate to an
existence theorem for rough initial data requires the existence of a
sequence of classical solutions of the constraints approximating the
rough solution; this approximation theorem was also missing.

In this paper we construct a family of solutions in $H^s\loc\times
H^{s-1}\loc$ with $s>3/2$.  We prove moreover that every solution in this
family can be approximated by a sequence of classical solutions.  To
compare the lower bound $s>3/2$ with previous results, we must keep in
mind that the constraint equations have typically been solved either
in H\"older spaces or in Sobolev spaces $\W {k}p\loc\times\W {k-1}p\loc$,
where $k$ is an integer.  The classical lower bound
\cite{const-asympt-flat} for the existence of solutions of the
constraint equations was $k>3/p+1$.  These metrics have one continuous
derivative and can be thought of as analogous to metrics in $H^s\loc$
with $s>5/2$.  This lower bound was improved in the settings of compact
manifolds \cite{cb-low-reg} and asymptotically Euclidean manifolds
\cite{maxwell-ahc} to $k\ge2$ and $k>3/p$. The restriction $k>3/p$ 
ensures that the metric is continuous, while the inequality $k\ge 2$
further implies that the curvature belongs to $L^p\loc$.  Taking $k=2$ and
$p=2$, these results provide for $H^2\loc\times H^1\loc$ solutions of the
constraint equations.  But they do not construct solutions directly in
the spaces of interest ($H^s\loc\times H^{s-1}\loc$ with $s>2$), nor do they
provide an approximation theorem.

Since $s=3/2$ is the scaling limit for the Einstein equations, this is
a natural lower bound for local well-posedness results.  It has been
suggested \cite{causal-geom-rough} that it might not be possible to
obtain local well-posedness down to $s>3/2$.  In the case of the
constraint equations, however, we have shown here that working in
spaces with $s>3/2$ is feasible.  In fact, the restriction $s>3/2$ is
analogous to the condition $k>3/p$ from \cite{cb-low-reg}
\cite{maxwell-ahc} as these thresholds ensure the metric
is continuous.  A novel feature of the solutions considered here
is that when $3/2 < s < 2$, the curvature of $g$ is in general only
a distribution, not necessarily an integrable function.

We restrict our attention to maximal (i.e. $\tr K=0$), asymptotically 
Euclidean (AE) solutions.  In the classical setting, the the conformal
method  of Lichnerowicz  \cite{lich-conformal}, Choquet-Bruhat and 
York \cite{cb-york-held} provides a parameterization of all such
solutions.  We review this construction below, as it forms the basis
of our construction of rough solutions.

Hereafter we suppose $M$ is an $n$-manifold with $n\ge 3$.
We seek a solution of the form
\bea
\hat g = \phi^{4\over n-2} g \\
\hat K = \phi^{-2} \sigma, \\
\eea
where $(M,g)$ is asymptotically Euclidean, $\sigma$ is a trace-free 
symmetric $(0,2)$-tensor, and $\phi$ is a conformal factor tending to 1 at 
infinity.  Writing the
constraint equations for $\hat g$ and $\hat K$ in terms of the conformal
data $(g,\sigma)$ we obtain
\bean
-a\Lap \phi + R\phi -\abs{\sigma}^2\phi^{-2\dimc-3}&= 0
\label{conformal-h}\\
\div \sigma &= 0,\label{conformal-m}\\
\eean
where $\dimc={2\over n-2}$ and $a=2\dimc+4$ are dimensional constants.
These equations for $\phi$ and $\sigma$ are decoupled (this is a feature 
of the hypothesis $\tr K=0$) and can be therefore be treated separately.

To construct solutions of \eqref{conformal-m}, known as
{\it transverse-traceless} tensors, we consider the conformal Killing
operator $\ck$ and vector Laplacian $\vl$.  These are defined by
\bea
\ck X & = \calL_X g-{2\over n}\div X g\\
\vl X & = \div\ck X,\\
\eea
where $\calL_X$ is the Lie derivative with respect to $X$.
If $S$ is a symmetric, traceless $(0,2)$-tensor, and if we can solve
\be
\vl X = -\div S,
\ee
then setting $\sigma = S+\ck X$ we have $\div \sigma = 0$.  The fact
that this is always possible, and that the set $\calS$ of
symmetric, traceless, $(0,2)$-tensors can be written
$\calS = \im\ck \oplus \ker\div$ is known as the York decomposition.

The study the Lichnerowicz equation \eqref{conformal-h},
now reduces to determining the set of metrics $(M,g)$ and transverse-traceless 
tensors $\sigma$ for which \eqref{conformal-h} admits a (positive) solution.
It was shown by Cantor in \cite{cantor-ae} that \eqref{conformal-h} is solvable if and only
if $(M,g)$ is conformally related to a scalar flat metric.
Following \cite{maxwell-ahc}, we define the
Yamabe invariant $\lambda_g$ of $(M,g)$ to be
\be
\lambda_{g}=\inf_{{f\in C\cpct^{\infty}(M)\atop f\not\equiv0}} {\int_M a \abs{\nabla f}^2 + R f^2\dV \over 
|| f ||_{L^{2n\over n-2}}^2}.
\ee
In \cite{lap-asympt-flat}, Cantor and Brill considered a related
invariant in their investigation of necessary and sufficient conditions
for an asymptotically Euclidean metric to be conformally related to
a scalar flat metric.  It was
shown in \cite{maxwell-ahc} that $(M,g)$ is conformally related to a
scalar flat metric if and only if
$\lambda_g>0$.  This provides a necessary and sufficient condition
for the solvability of \eqref{conformal-h}, which we will call
the Cantor-Brill condition.  Combining the York decomposition and the
Cantor-Brill condition, we obtain an elegant construction of the
entire family of maximal AE solutions of the constraint equations.

The main results of this paper are that the York decomposition
and the Cantor-Brill condition are valid for rough metrics (i.e. metrics with
$s>n/2$), and that rough,
maximal, AE solutions of the constraints admit a sequence of classical
approximating solutions.  At the core of each of these problems is 
a second-order linear elliptic operator with rough coefficients.  
For the York decomposition we have the vector
Laplacian $\vl$, and for the Cantor-Brill condition we have the
linearized Lichnerowicz equation
\be
-a\Lap + V,
\ee
where $V$ is a potential.  Both operators have the structure
\be
a_2\partial^2+a_1\partial+a_0
\ee
where $a_i\in H^{s-2+i}\loc$.  In Section \ref{sec:elliptic} we obtain
mapping properties and a-priori estimates for operators with these low
regularity coefficients.  Once the a-priori estimates have been
achieved, the York decomposition follows in a straight-forward way in Section
\ref{sec:york}.  We also show in section \ref{sec:york} that there are
no conformal Killing fields decaying at infinity in $H^s$ with
$s>n/2$.  This fact extends a corresponding result for classical
metrics \cite{boost-problem} and is used crucially in Section
\ref{sec:approx} to prove the existence of classical approximating
sequences. Section \ref{sec:linearlich} obtains properties
of the linearized Lichnerowicz equation, and Section \ref{sec:lich} applies 
these results to solve the Lichnerowicz equation given the Cantor-Brill 
property.

In the discussion above, we have suppressed important technical details
about the function spaces which we use. The maximal, asymptotically Euclidean
setting has the advantage of possessing a very clean theory, but also
requires the use of weighted $H^s$ spaces to prescribe asymptotic
behaviour near infinity.  These spaces are routinely used in the
mathematical relativity community in the special case where $s$ is a
non-negative integer.  There is a generalized version of these spaces
for $s\in\Reals$ due to Triebel
\cite{triebel-weighted-one}\cite{triebel-weighted-two} which
has not been exploited before to solve the constraint equations.
In the following section we define these spaces and prove some simple but 
fundamental facts about multiplying functions in these spaces.

\section{Weighted Sobolev Spaces}
\seclabel{sec:fs}

Let $A_j$ be the annulus $B_{2^{j+1}}\setminus 
\overline{B_{2^{j-1}}}$, and let $\{\phi_j \}_{j=0}^{\infty}$
be a partition of unity satisfying
\bl
\li{1.} $\phi_0$ is supported in $B_2$ and
$\phi_j$ is supported in $A_j$ for $j\ge 1$. 
\li{2.} $\phi_j(x) = \phi_1(2^{1-j} x)$ for $j\ge 1$.

\el
The weighted Sobolev space $H^s_\delta(\Rn)$ with $s\in\Reals$ and 
$\delta\in\Reals$ is then defined as a subset of the tempered 
distributions $\calS^*$,
\be
H^s_\delta(\Rn) = \left\{ u\in \calS^* : 
||u||_{H^s_\delta}^2=\sum_{j=0}^\infty 2^{-2\delta j}
|| T_j(\phi_j u) ||_{H^s}^2  < \infty\right\}.
\ee
Here $T_j$ is the rescaling operator defined by $(T_ju)(x) = u(2^j x)$.
If $\Omega$ is a open subset of $\Rn$, we define 
$H^s_\delta(\Omega)$ to be the restriction of $H^s_\delta(\Rn)$ to 
$\Omega$ with norm
\be
||u||_{H^s_\delta(\Omega)}=\inf_{v\in H^s_\delta(\Rn)\atop \left.v\right|_\Omega =u} 
||v||_{H^s_\delta(\Rn)}.
\ee

For properties of these spaces we refer the reader to 
\cite{triebel-weighted-one}\cite{triebel-weighted-two}, where our spaces 
$H^s_\delta$ correspond with 
the spaces $h^s_{2,2s-2\delta-n}$ in that reference.  In particular, 
$H^s_\delta$ is a reflexive Banach space, and the norms corresponding to 
different partitions of unity are all equivalent.  We use Bartnik's 
convention \cite{mass-asympt-flat} for the growth parameter $\delta$
so that functions in $H^s_\delta$ grow no faster than $\abs{x}^\delta$
near infinity.  
These spaces are related to the more familiar spaces $\W kp\del$ defined 
for $k$ a non-negative integer and $p\in(1,\infty)$ by
\be
\W kp_\delta(\Rn) = \left\{ u\in \calS^* : 
||u||_{\W kp_\delta(\Rn)}=
\sum_{|\beta|\le k} ||(1+\abs{x})^{-\delta-\expfrac np+|\beta|}\,
\partial_\beta u ||_{L^p(\Rn)}< \infty\right\}.
\ee
We have the following basic facts.

\lem\label{lemma:basicweighted}
\bl
\li{1.} When $k$ is a non-negative integer, an equivalent norm
for $H^k_\delta$ is the norm on $\W k2\del$.
\li{2.}  If $\theta\in(0,1)$, $s=(1-\theta)s_1+\theta s_2$ and 
$\delta = (1-\theta)\delta_1 +\theta\delta_2$, then 
$H^s_\delta$ is the interpolation space 
$[H^{s_1}_{\delta_1},H^{s_2}_{\delta_2}]_\theta$.
\li{3.} $H^s_\delta$ is the dual space of $H^{-s}_{-n-\delta}$.
\li{4.} If $s\ge s'$ and $\delta\le \delta'$ then 
$H^{s}_{\delta}$ is continuously embedded in $H^{s'}_{\delta'}$ .  If $s>s'$ 
and $\delta<\delta'$, then this embedding is compact.
\li{5.} If $s<n/2$, then $H^s\del$ is continuously embedded in 
$L^q\del=\W 0q\del$, where ${1\over q} = {1\over 2} - {s\over n}$.
\li{6.} If $s>n/2$ then $H^s_\delta(\Rn)$ is continuously embedded in
$C^0_\delta(\Rn)$, where 
$||f||_{C^0_\delta} = \sup_{x\in\Rn} (1+\abs{x})^{-\delta}\abs{f}.$
\li{7.} If $u\in H^s\del$, then $\partial u \in H^{s-1}_{\delta-1}$.

\el
\Proof
All these claims follow immediately from \cite{triebel-weighted-one}
\cite{triebel-weighted-two} except for claims 6, 7 and the compact embedding
$H^{s}_{\delta}\hookrightarrow H^{s'}_{\delta'}$ when $s>s'$ and 
$\delta<\delta'$.

To prove claim 6, we know from \cite{triebel-weighted-one} that 
$H^s_\delta$ embedded in $\W kp_\delta$ for $k$ an integer with $k<s$ and 
$k>n/p$.  The claim now follows from the corresponding fact for $\W kp_\delta$
spaces.  Claim 7 is well known for $H^s_\delta$ with $s$ an integer
and $s\ge 1$.  The result for $s\in\Reals$ follows easily from
duality and interpolation.

We now turn to the compactness argument.
Let $\{u_k\}_{k=1}^\infty$ be a sequence in $H^s\del$ 
with $||u_k||_{H^s\del}\le1$.   Then the sequence 
$\{\phi_0 u_k\}_{k=1}^\infty$ is bounded in $H^s$ and each element has
support contained in the ball $B_2$.  From Rellich's theorem, we
infer the existence of a subsequence $\{u_k^0\}_{k=0}^\infty$ such that
$\{\phi_0 u^0_k \}_{k=1}^\infty$ is 
Cauchy in $H^{s'}$ and such that $||\chi_0(u^0_k-u^0_l)||_{H^{s'}}\le 1$ for
all $k,l\ge 1$.  Similarly , the sequence $\{T_1(\phi_1 u^0_k)\}_{k=1}^\infty$
is bounded in $H^s$ and has uniformly compact support.  So there is 
a sub-subsequence $\{u^1_k\}_{k=1}^\infty$ such that 
$\{T_1(\phi_1 u^1_k)\}_{k=1}^\infty$ is Cauchy in $H^{s'}$ and
\be
\sum_{j=0}^1 2^{-2\delta'j}||T_j(\phi_j (u^1_k-u^1_l)||_{H^{s'}}^2 < {1\over 2}
\ee
for $k,l \ge 1$.  Continuing iteratively, we obtain sub-subsequences 
$\{u^m_k\}_{k=1}^\infty$ such that for $k,l\ge 1$,
\be
\sum_{j=0}^m 2^{-2\delta'j}||T_j(\phi_j (u^m_k-u^m_l)||_{H^{s'}}^2 < {1\over 2^m}
\ee
Let $v_k=u^k_1$.  Then if $k,l\ge N$, since $\delta'>\delta$,
\bea
||v_k-v_l||_{H^{s'}_{\delta'}}^2 &= 
\sum_{j=0}^\infty 2^{-2\delta'j}||T_j(\phi_j(v_k-v_l))||_{H^{s'}}^2\\
& = \sum_{j=0}^N 2^{-2\delta'j}||T_j(\phi_j(v_k-v_l))||_{H^{s'}}^2+
\sum_{j=N+1}^\infty 2^{-2\delta'j}||T_j(\phi_j(v_k-v_l))||_{H^{s'}}^2\\
&\le 2^{-N} + 2^{-2(\delta'-\delta)(N+1)}
\sum_{j=N+1}^\infty 2^{-2\delta j}||T_j(\phi_j(v_k-v_l))||_{H^{s}}^2\\
&\le 2^{-N} + 2^{-2(\delta'-\delta)(N+1)}|| v_k-v_l||_{H^s\del}^2.
\eea
Since the sequence $\{v_k\}_{k=1}^\infty$ is bounded in $H^s_\delta$ and
since $\delta'>\delta$, we conclude the sequence is Cauchy in 
$H^{s'}_{\delta'}$, which proves the result.
\EOProof

We define asymptotically Euclidean manifolds in the usual way using weighted
spaces.
\Definition
Let $M$ be a smooth, connected, $n$-dimensional manifold and let $g$ be a 
metric on $M$ for which $(M,g)$ is complete.  Let $E_r$ be
the exterior region $\{x\in \Rn:\abs{x}>r\}$.  For $s>n/2$ and $\rho<0$,
we say $(M,g)$ is asymptotically Euclidean (AE) of class $H^s\srho$ if 
\bl
\li{1.} The metric $g\in H^s\loc(M)$.
\li{2.} There exists a finite collection $\{N_i\}_{i=1}^m$ of open subsets
of $M$ and diffeomorphisms $\Phi_i:E_1\mapsto N_i$ such that $M-\cup_i N_i$
is compact.
\li{3.} For each $i$, $\Phi_i^*g-\gflat\in H^s\srho(E_1)$, where 
$\gflat$ is the Euclidean metric.
\el

\EODefinition

The charts $\Phi_i$ are called end charts and the corresponding coordinates
are end coordinates. Suppose $(M,g)$ is asymptotically Euclidean, and 
let $\{\Phi_i\}_{i=1}^m$ be its collection of end charts.  
Let $U_0 = M - \overline{\cup_i \Phi_i( E_2)}$, and let $\{\chi_i\}_{i=0}^m$
be a partition of unity subordinate to the sets $\{U_0,N_1,\cdots,N_n\}$.
The weighted Sobolev space $H^s_\delta(M)$ is the subset of 
$H^s\loc(M)$ such that the
norm
\be
||u||_{H^s_\delta(M)} = ||\chi_0 u||_{H^s(U_0)} + 
\sum_{i=1}^m||\Phi_i^*(\chi_i u)||_{H^s\del(\Rn)}
\ee
is finite.  
An initial data set $(M,g,K)$ is asymptotically Euclidean if $(M,g)$ is
asymptotically Euclidean of class $H^s\srho$ for $s>n/2$ and $\rho<0$,
and if $K\in H^{s-1}_{\rho-1}(M)$.  We note that when $n=3$ and $(M,g,K)$ 
is AE of class $H^s_\rho$ with $s>2$ and $\rho\le -1/2$, then 
$(\nabla g, K)\in H^{s-1}\times H^{s-1}$ and hence $(M,g,K)$ is 
initial data that can be used, for example, in the theorems 
of \cite{kr-rough-annals}.

The remainder of this section concerns multiplication and Sobolev
spaces; we need these properties to work with differential operators with
rough coefficients. We recall the following multiplication rule for
unweighted Sobolev spaces, which is an easy consequence of
the well-known fact that $H^s$ is an algebra for $s>n/2$ together
with interpolation and duality arguments.

\lem\label{lemma:basicmult}
Suppose $t \le \min (s_1,s_2)$, $s_1+s_2\ge 0$ and $t < s_1+s_2-{n\over 2}$.
Then pointwise multiplication extends to a continuous bilinear map
\be
H^{s_1}\times 
H^{s_2}\rightarrow 
H^{t}.
\ee
\NoProof

The previous lemma generalizes to weighted spaces. In the following,
we use the notation $A\lesssim B$ to mean $A< c B$ for a certain
positive constant $c$.  The implicit constant is independent of the
functions and parameters appearing in $A$ and $B$ that are not assumed
to have a fixed value.

\lem\label{lemma:mult}
Suppose $s \le \min (s_1,s_2)$, $s_1+s_2\ge 0$, and $s < s_1+s_2-{n\over 2}$.
For any $\delta_1,\delta_2\in\Reals$, pointwise multiplication extends
to a continuous bilinear map
\be
H^{s_1}_{\delta_1}\times 
H^{s_2}_{\delta_2}\rightarrow 
H^{s}_{\delta_1+\delta_2}.
\ee
\Proof
Suppose $u_i\in H^{s_i}_{\delta_i}$.  Taking $\phi_k=0$ for $k<0$ we have 
\be
T_j(\phi_j u_1 u_2) = T_j(\phi_j) \sum_{k=j-1}^{j+1} 
T_j(\phi_k u_1) \sum_{l=j-1}^{j+1} T_j(\phi_l u_2).
\ee
From the restrictions on $s$, $s_1$, and $s_2$ we know that 
multiplication is a continuous bilinear map on the corresponding unweighted 
Sobolev spaces.  Noting that $T_j\phi_j=T_k\phi_k$ for $j,k\ge 1$, 
we find
\bea
||T_j( \phi_j u_1 u_2 )||_{H^s}^2 &\lesssim
\sum_{k,l=j-1}^{j+1} 
||T_j(\phi_k u_1)||_{H^{s_1}}^2 
||T_j(\phi_l u_2)||_{H^{s_2}}^2 \\
&\lesssim 
\sum_{k,l=j-1}^{j+1}  
||T_{j-k}T_k(\phi_k u_1)||_{H^{s_1}}^2 
||T_{j-l}T_l(\phi_l u_2)||_{H^{s_2}}^2.
\eea
Now $T_{j-k}$ must be one of $T_{-1}$, $T_0$, or $T_1$, and a same result
holds for $T_{j-l}$.  These operators are independent of $j$ and we find
\be
||T_j( \phi_j u_1 u_2 )||_{H^s}^2 \lesssim
\sum_{k,l=j-1}^{j+1}  
||T_k(\phi_k u_1)||_{H^{s_1}}^2 
||T_l(\phi_l u_2)||_{H^{s_2}}^2.
\ee
It follows that
\bea
\sum_{j=0}^{\infty} 2^{-2\delta j} ||T_j(\phi_j u_1 u_2)||_{H^s}^2
&\lesssim
\sum_{j=0}^{\infty} 2^{-2\delta j}
\sum_{k,l=j-1}^{j+1} 
||T_k(\phi_k u_1)||_{H^{s_1}}^2 
||T_l(\phi_l u_2)||_{H^{s_2}}^2\\
&\lesssim
\sum_{k=-1}^{1}\sum_{j=0}^{\infty}
\left[2^{-2\delta_1 j}||T_j(\phi_j u_1)||_{H^{s_1}}^2\times\right.\\
&\qquad\qquad\qquad\qquad\qquad
\times 
\left.2^{-2\delta_2(j+k)} 
||T_{j+k}(\phi_{j+k} u_2)||_{H^{s_2}}^2\right]\\
&\lesssim
\sum_{j=0}^{\infty} 2^{-2\delta_1 j}||T_j(\phi_j u_1)||_{H^{s_1}}^2 
\sum_{k=0}^{\infty} 2^{-2\delta_2 k}||T_k(\phi_k u_2)||_{H^{s_2}}^2.
\eea
This proves 
$||u_1 u_2||_{H^s_{\delta_1+\delta_2}}\lesssim 
||u_1||_{H^{s_1}_{\delta_1}} ||u_2||_{H^{s_2}_{\delta_2}}$.
\EOProof

The following results, related to the multiplication lemmas,
are useful for working with nonlinearities.

\lem\label{lem:nonlin}
Suppose $f:\Reals\ra\Reals$ is smooth. If
$u\in H^s_\rho$ with $s>n/2$ and $\rho < 0$, and  
if $v\in H^\sigma_{\delta}$ with $\sigma\in[-s,s]$ and $\delta\in \Reals$, then 
\be
f(u)v \in H^\sigma_\delta.
\ee
Moreover, the map taking $(u,v)$ to $f(u)v$ is continuous 
from $H^s_\rho\times H^\sigma_\delta$ to $H^\sigma_\delta$.
\Proof
It is easy to verify with the help of Lemma \ref{lemma:mult} that if 
$u\in H^s$ with $s>n/2$, and if $\eta$ is smooth and compactly supported, 
then $\eta f(u)\in H^s$ and the map taking $u$ to $\eta f(u)$ is continuous 
from $H^s$ to $H^s$. 

Now suppose $u$ and $v$ satisfy the hypotheses of the lemma.  Then
\bea
||f(u)v ||_{H^\sigma_\delta}^2 &=
\sum_{j=0}^\infty 2^{-2\delta j} || T_j(\phi_j f(u)v) ||_{H^\sigma}^2 \\
&= \sum_{j=0}^\infty 2^{-2\delta j} 
\bnorm{\sum_{k=j-1}^{j+1} (T_j \phi_k) f\left( T_j \sum_{i=j-1}^{j+1}\phi_i u 
\right) T_j \phi_j v }_{H^\sigma}^2 \\
&\lesssim \sum_{j=0}^\infty 2^{-2\delta j} 
\bnorm{ \sum_{k=j-1}^{j+1} (T_j \phi_k) f\left( R_j u\right)}_{H^s}^2
||T_j \phi_j v ||_{H^\sigma}^2,
\eea
where $R_j u = \sum_{i=j-1}^{j+1}T_{j-i} T_i \phi_i u$

Let $\eta=\sum _{k=0}^2 T_k \phi_1$, so that for $j>1$ we
have $\sum_{k=j-1}^{j+1} (T_j \phi_k)=\eta$.  Since $\rho>0$, $T_i\phi_i u$ 
converges to 0 in $H^s$.  It follows that 
$R_j u $ converges to $0$ in $H^s$ 
as well. Hence $\eta f\left(R_j u \right)$ converges in
$H^s$ to $\eta f(0)$, and  we conclude that there exists a bound $M$ such that
\bel{nonlin-inter1}
\bnorm{\eta (T_j \phi_k) f\left( R_j u\right) }_{H^s}\le M
\eel
for all $j>1$.  The cases $j=0$ and $j=1$ can be treated similarly.
We conclude, taking $M$ sufficiently large, that 
\bel{nonlin-inter2}
||f(u)v ||_{H^\sigma_\delta}^2\lesssim M^2 ||v||_{H^\sigma_\delta}^2.
\eel
This proves $f(u)v\in H^\sigma_\delta$.

To establish the continuity of the map $(u,v)\mapsto f(u)v$ acting on
$H^s_\rho \times H^\sigma_\delta$, we consider any sequence 
$\{u_k,v_k\}_{k=1}^\infty$ converging to $(u,v)$.  Then 
\be
f(u)v-f(u_k)v_k =  f(u)(v-v_k)-(f(u)-f(u_k))v_k.
\ee
From \eqref{nonlin-inter2} we see that $f(u)(v-v_k)\ra 0$ in $H^\sigma_\delta$.
We wish to establish $(f(u)-f(u_k))v_k\ra 0$ as well.  

Computing as before we find
\bea
||(f(u)-f(u_k))v_k||_{H^\sigma_\delta}^2&\lesssim \sum_{j=0}^\infty 2^{-2\delta j} 
||T_j\phi_j v_k||_{H^\sigma}^2
\bnorm{\sum_{l=j-1}^{j+1} (T_j\phi_l) \left[ f\left(R_j u\right) 
- f\left(R_j u_k\right) \right]}_{H^s}^2\\
&\lesssim ||v_k||_{H^\sigma_\delta} \, \sup_{j\ge 0}\;
\bnorm{\sum_{l=j-1}^{j+1} (T_j\phi_l) \left[ f\left(R_j u\right) 
- f\left(R_j u_k\right) \right]}_{H^s}^2.
\eea
 Since $\{v_k\}_{k=1}^\infty$ is bounded in $H^\sigma_\delta$, it is enough to 
show
\bel{nonlin-converge}
\sup_{j\ge 0}\;
\bnorm{\sum_{l=j-1}^{j+1} (T_j\phi_l) \left[ f\left( R_j u \right) 
- f\left(R_j u_k\right) \right]}_{H^s}^2\ra 0
\eel
as $k\ra\infty$.

Since
$||T_j\phi_j(w)||_{H^s} \le 2^{\delta j}||w||_{H^s_\delta}$ 
for all $w\in H^s_\delta$, we also have
$|| R_j w ||_{H^s} \lesssim 2^{\delta j} ||w||_{H^s_\delta}$.
Hence
\beal{Rjbounds}
|| R_j u ||_{H^s} &\lesssim 2^{\delta j} ||u||_{H^s_\delta} \\
|| R_j u_k||_{H^s} &\lesssim  2^{\delta j} 
\left( ||u||_{H^s_\delta}+\sup_{k\ge 1} ||u_k||_{H^s_\delta}\right)
\eeal
Now
\bel{Rjsplit}
||\eta \left[ f( R_j u) - f(R_j u_k)\right]||_{H^s}
\le
||\eta f( R_j u) - \eta f(0) ||_{H^s} + ||\eta f(R_j u_k)-\eta f(0)||_{H^s}
\eel
where, as before, $\eta=\sum _{k=0}^2 T_k \phi_1=
\sum_{l=j-1}^{j+1} (T_j \phi_l)$ for $j>1$.  
Since $\delta <0$ and since the map $u\mapsto \eta f(u)$ is continuous from 
$H^s$ to $H^s$, we conclude from \eqref{Rjbounds} and \eqref{Rjsplit} that
\be
\sup_{j> N} \sup_{k\ge 1}\;
||\eta \left[ f( R_j u) - f(R_j u_k)\right]||_{H^s}
\ee
can be made arbitrarily small for $N$ sufficiently large.
On the other hand, for any fixed $N$
\be
\sup_{j= 0}^{N}\;
\bnorm{\sum_{l=j-1}^{j+1} (T_j\phi_l) \left[ f\left(R_j u\right) 
- f\left(R_j u_k\right) \right]}_{H^s}
\ee
can be made as small as we please by taking $k$ sufficiently large, since each
of the finitely many terms in the supremum tends to 0 as $k$ goes to infinity.
We have hence established \eqref{nonlin-converge} and therefore also the 
desired continuity.
\EOProof
To work with conformal changes of an AE metric, we need a variation
of Lemma \ref{lem:nonlin}.  If $g$ is AE of class $H^s_\rho$,
and if $f$ is smooth with $f(0)=1$, we want to 
know that $f(u)g$ is AE of class $H^s_\rho$ whenever $u\in H^s_\rho$.  
Now
\be
f(u)g - \gflat  =  f(u)(g-\gflat) + (f(u)-1)\gflat.
\ee
Since $g-\gflat\in H^s_\rho$, we know from Lemma \ref{lem:nonlin} that
$f(u)(g-\gflat)\in H^s_\rho$.  The following corollary shows that
$(f(u)-1)\gflat$ belongs to $H^s_\rho$ as well.  
\cor\label{cor:nonlin}
Suppose $f:\Reals\ra\Reals$ is smooth and $f(0)=0$. If $u\in H^s_\rho$
with $s>n/2$ and $\rho<0$ then $f(u)\in H^s_\rho$ and the map taking
$u$ to $f(u)$ is continuous from $H^s_\rho$ to $H^s_\rho$.
\Proof
Since $f(0)=0$ we have from Taylor's theorem that $f(x)=F(x)x$ where
$F$ is smooth.  Hence $f(u)=F(u)u\in H^s_\rho$ by Lemma \ref{lem:nonlin},
and the continuity of the map on $H^s_\rho$ follows similarly.
\EOProof
\Remark  In practice we will use an obvious improvement to 
Lemma \ref{lem:nonlin} and Corollary \ref{cor:nonlin}. It is
easy to see that if $f$ is only smooth on an open interval $I$, and if 
$[\inf u, \sup u]\subset I$, 
then $f(u)v\in H^\sigma_\delta$ and the map
$(u,v)\mapsto f(u)v$ is continuous on $U\times H^\sigma_\delta$ for
some neighbourhood $U$ of $u$. An analogous statement for Corollary
\ref{cor:nonlin} also holds.
\EORemark

For functions $u,v\in C^k$ we have a simple estimate on how the
$k^{\rm th}$ derivatives of $v$ contribute to the $C^k$ norm of $uv$,
namely $||uv||_{C^k}\lesssim 
||u||_{C^0}||v||_{C^k} + ||u||_{C^{k}}||v||_{C^{k-1}}$.  We need an
analogous fact for Sobolev spaces, which will be derived from the 
following commutator estimate for the operator $\Lambda=(1-\Lap)^{1/2}$.

\lem\label{lemma:commutator}
Suppose $u\in H^s$ with $s>n/2$, $\sigma\in(-s,s)$, and 
suppose $\tau\in(0,1]$ satisfies $\tau< \min(s-n/2, s+\sigma)$.  Then
$[\Lambda^{-\sigma},u]$ is continuous as a map
\be
[\Lambda^{-\sigma},u]: H^{-\sigma}\rightarrow H^\tau.
\ee
\Proof
Since $\Lambda^\sigma$ is a pseudo-differential operator of order $\sigma$,
and since $u$ is a pseudo-differential operator of order 0 of the type
considered in \cite{marschall}, the claim is a consequence
of \cite{marschall} Corollary 3.4.
\EOProof

\lem\label{lemma:tau}
Suppose $s>n/2$, $\sigma\in(-s,s]$,
and $\tau\in(0,1]$ satisfies $\tau< \min(s-n/2, s+\sigma)$. Then 
for all $u\in H^s$ and $v\in H^\sigma$ there is a constant $c(u)$ such 
that
\bel{small-prod-est}
||u v||_{H^\sigma} \lesssim ||u||_{L^\infty} ||v||_{H^\sigma} + 
c(u) ||v||_{H^{\sigma-\tau}}.
\eel
\Proof
First suppose $\sigma\in(-s,s)$.  Then for all 
$\phi\in C^\infty\cpct$,
\bea
\ip<uv,\phi> & = \ip<\Lambda^{\sigma} v, \Lambda^{-\sigma} u \phi> \\
& =  \ip< \Lambda^{\sigma} v, u \Lambda^{-\sigma} \phi> + 
\ip< v, \Lambda^{\sigma} [ \Lambda^{-\sigma},u ] \phi>. \\
&\lesssim 
||u||_{L^\infty}||v||_{H^\sigma}||\phi||_{H^{-\sigma}}
+
||v||_{H^{\sigma-\tau}}||[ \Lambda^{-\sigma},u ] \phi||_{H^\tau}.
\eea
From Lemma \ref{lemma:commutator}, 
\be
||[ \Lambda^{-\sigma},u ] \phi||_{H^\tau}\lesssim
c(u) ||\phi||_{H^{-\sigma}}.
\ee
Hence
\bel{prodest}
||uv||_{H^\sigma}\lesssim ||u||_{L^\infty} ||v||_{H^\sigma}+
c(u) || v||_{H^{\sigma-\tau}}.
\eel
In the case $\sigma=s$, we have from \cite{katoponce}
\bea
||uv||_{H^\sigma}
&\lesssim ||u||_{L^\infty} ||v||_{H^s}+||u||_{H^s} || v||_{L^\infty}\\
&\lesssim ||u||_{L^\infty} ||v||_{H^\sigma}+||u||_{H^s} || v||_{H^{\sigma-\tau}},
\eea
which completes the proof.
\EOProof
In fact, one can show that $c(u)=||u||_{H^s}$ in the previous lemma, but we
we do not need this fact in the sequel.

\section{Elliptic Linear Operators with Rough Coefficients}
\seclabel{sec:elliptic}

The regularity of the coefficients of the vector Laplacian $\vl$ and 
the conformal Laplacian $-a\Lap+R$ depends on the regularity of the
metric $g$.  The following general class of differential operators 
$\calL^{m,s}_\rho$ encodes this relationship.

\Definition
Let $A$ be the linear differential operator
\be
A = \sum_{\abs{\alpha}\le m} a^\alpha \partial_\alpha,
\ee
where $a^\alpha$ is a $\Reals^{k\times k}$ valued function.
We say that 
\be
A\in \calL^{m,s}
\ee
if  $a^\alpha\in H^{s-m+\abs\alpha}$ for all $\abs{\alpha}\le m$. Similarly,
if $\rho< 0$ we say
\be
A\in\calL^{m,s}_{\rho}
\ee
if $a^\alpha\in H^{s-m+\abs\alpha}_{\rho-m+\abs\alpha}$ for all 
$\abs\alpha< m$ and if there are constant matrices $a^\alpha_\infty$ such that
$a^\alpha_\infty-a^\alpha\in H^{s}_{\rho}$ for all $\abs\alpha=m$.  We call
$A_\infty= \sum_{\abs\alpha = m} a^\alpha_\infty\partial_\alpha$ the principal
part of $A$ at infinity.
\EODefinition

One can easily check with the help of Lemma \ref{lem:nonlin} and Corollary
\ref{cor:nonlin} that when $g$ is an AE metric of class $H^s_\rho$ with
$s>n/2$ and $\rho<0$, then 
both the vector Laplacian and the conformal Laplacian are in 
$\calL^{2,s}_\rho$.
From Lemmas \ref{lemma:basicmult} and \ref{lemma:mult} we obtain the following 
simple properties.
\cor\label{cor:opmap}
Suppose 
\bea
\eta&<\sigma+s-m-{n\over 2} \\
\eta&\le \min(\sigma,s)-m\\
m&\le \sigma+s.
\eea
If $A\in\calL^{m,s}$, then $A$ is a continuous map
\be
A:H^{\sigma}\rightarrow H^{\eta}.
\ee
If $\delta\in\Reals$, $\rho<0$ and if  $A\in\calL^{m,s}_{\rho}$, then $A$ 
is a continuous map
\be
A:H^{\sigma}_{\delta}\rightarrow H^{\eta}_{\delta-m}.
\ee
If moreover $A_\infty=0$, then $A$ is a continuous map
\be
A:H^{\sigma}_{\delta}\rightarrow H^{\eta}_{\delta-m+\rho}.
\ee
\NoProof

When $s>n/2$, the highest order coefficients of $A\in\calL^{m,s}$ are 
continuous. It then makes sense to talk about their pointwise values.
We say $A$ is elliptic if for each $x$, the constant coefficient operator
$\sum_{\abs\alpha=m}a^\alpha(x)\partial_\alpha$ is elliptic.  
For $A\in\calL^{m,s}_\rho$
we require additionally that $A_\infty$ is elliptic. 

We now prove an
interior regularity a-priori estimate for elliptic operators in
$\calL^{m,s}$ in two steps.  We first we treat the high order terms
using coefficient freezing and Lemma \ref{lemma:tau}. 

\prop\label{prop:elliptic-local-high}
Suppose $s>n/2$ and suppose $A\in\calL^{m,s}$ is elliptic
and has only has terms of order $m$ (i.e.
$A=\sum_{\abs\alpha=m}a^\alpha(x)\partial_\alpha$).
If $\sigma\in(m-s,s]$, then for all
$u\in H^\sigma$ supported in a compact set $K$
\bel{elliptic-local-high-est}
||u||_{\H \sigma} \lesssim || A u ||_{H^{\sigma-m}} + 
||u||_{H^{m-s}},
\eel
where the implicit constant depends on $K$ and $A$ but not on $u$.
\Proof
Fix $x_0\in K$ and let $\epsilon$ be a small parameter to be chosen
later. Let $\chi$ be a cutoff function equal to 1 on $B_1$ and equal to
0 outside $B_2$, and let $\chi_\epsilon(x)=\chi((x-x_0)/\epsilon)$.
Let $A=A_m+R$ where $A_m$ is the constant coefficient operator
$\sum_{\abs\alpha=m} a^\alpha(x_0)\partial_\alpha$ and
$R=\sum_{\abs\alpha=m} r^\alpha\partial_\alpha=
\sum_{\abs\alpha=m} \left(a^\alpha-a^\alpha(x_0)\right)\partial_\alpha$.

From elliptic theory for constant coefficient elliptic operators we have
\bea
||\chi_\epsilon u||_{H^\sigma}&\lesssim || A_m\chi_\epsilon u||_{H^{\sigma-m}}
+||\chi_\epsilon u||_{H^{m-s}}\\
&\lesssim ||A\chi_\epsilon u||_{H^{\sigma-m}} + ||R\chi_\epsilon u||_{H^{\sigma-m}}
+||\chi_\epsilon u||_{H^{m-s}}\\
&\lesssim c(\epsilon)|| A u||_{H^{\sigma-m}} + 
|| [A,\chi_\epsilon] u||_{H^{\sigma-m}} + 
||R\chi_\epsilon u||_{H^{\sigma-m}}
+c(\epsilon)||u||_{H^{m-s}}.
\eea
To estimate the term $||R\chi_\epsilon u||_{H^{\sigma-m}}$, we have
\bea
||r^\alpha\partial_\alpha \chi_\epsilon u||_{H^{\sigma-m}}
& = 
||\chi_{2\epsilon} r^\alpha\partial_\alpha \chi_\epsilon u||_{H^{\sigma-m}}\\
&\lesssim 
||\chi_{2\epsilon}r^\alpha||_{L^\infty}||\chi_\epsilon u||_{H^\sigma}
+c(\chi_{2\epsilon} r^\alpha)||\chi_\epsilon u||_{H^{\sigma-\tau}},
\eea
where $\tau>0$ is a constant given by Lemma \ref{lemma:tau} satisfying
$\sigma-\tau > m-s$.  Taking $\epsilon$ sufficiently small we 
we can make $||\chi_{2\epsilon}r^\alpha||_{L^\infty}$ as small as we please
and we obtain for an $\epsilon$ depending only on $A$ and $x_0$
\be
||\chi_\epsilon u||_{H^\sigma} \lesssim 
c(\epsilon)|| Au||_{H^{\sigma-m}}
+ || [A,\chi_\epsilon] u||_{H^{\sigma-m}}
+ c(\epsilon, A,x_0)||u||_{H^{\sigma-\tau}}.
\ee
Now $[A,\chi_\epsilon]\in\calL^{m-1,s}$ and hence from Corollary
\ref{cor:opmap}
\be
|| [A,\chi_\epsilon] u||_{H^{\sigma-m}} \lesssim c(\epsilon)||u||_{H^{\sigma-1}}.
\ee
Since $\tau\le 1$, we obtain
\be
||u||_{H^\sigma({B_{\epsilon/2}(x)})} \lesssim 
c(\epsilon,A,x_0)\left[|| Au||_{H^{\sigma-m}}
+ ||u||_{H^{\sigma-\tau}}\right].
\ee
Covering $K$ with finitely many such balls we find
\be
||u||_{H^\sigma} \lesssim 
|| Au||_{H^{\sigma-m}}
+ ||u||_{H^{\sigma-\tau}},
\ee
where the implicit constant depends on $K$ and $A$.
Since $\sigma-\tau>m-s$, 
equation \eqref{elliptic-local-high-est} then follows from
interpolation.
\EOProof

We now can prove an interior regularity estimate for elliptic
operators in $\calL^{m,s}$.

\prop\label{prop:elliptic-local} 
Let $U$ and $V$ be open sets with $U\cpctsubset V$, and
suppose $s>n/2$ and  $\sigma\in(m-s,s]$.
If $A\in \calL^{m,s}$ is elliptic, then for every $u\in H^\sigma$ we have
\bel{elliptic-local-est}
||u||_{\H \sigma(U)} \lesssim || A u ||_{H^{\sigma-m}(V)} + 
||u||_{H^{m-s}(V)}.
\eel
\Proof
Choose an open set $V_0$ such that $U\cpctsubset V_0\cpctsubset V$,
and let $\chi$ be a cutoff function equal to 1 on $U$ and compactly
supported in $V_0$. Let $A=A_m+A\low$ where $A_m$ is the order $m$ operator
$\sum_{\abs\alpha=m} a^\alpha\partial_\alpha$.
From Proposition \ref{prop:elliptic-local-high} we have
\bean
||\chi u||_{H^\sigma}&\lesssim || A_m\chi u||_{H^{\sigma-m}}
+||\chi u||_{H^{m-s}}\nonumber\\
&\lesssim ||\chi Au||_{H^{\sigma-m}} + ||[A,\chi] u||_{H^{\sigma-m}}
+ ||A\low \chi u||_{H^{\sigma-m}}+||\chi u||_{H^{m-s}}.\label{local-one}
\eean
Let $\chi'$ be a second cutoff function equal
to 1 on $\supp \chi$ and also compactly supported in $V_0$.  Arguing
as in Proposition \ref{prop:elliptic-local-high} we have
\bel{local-two}
||[A,\chi] u||_{H^{\sigma-m}} \lesssim ||\chi' u||_{H^{\sigma-1}}.
\eel
Now $A\low\in\calL^{m-1,s-1}$.  Pick $\tau$ such that $\tau< s-{n\over 2}$
and $\tau\le\min(1, \sigma-(m-s))$.  Then from Corollary \ref{cor:opmap}
we have
\bel{local-three}
||A\low \chi u||_{H^{\sigma-m}}
\lesssim ||\chi u||_{H^{\sigma-\tau}}.
\eel
Combining equations \eqref{local-one}--\eqref{local-three}, we obtain
\be
||u||_{H^\sigma(U)}\lesssim
||Au||_{H^{\sigma-m}(V_0)} + ||u||_{H^{\sigma-\tau}(V_0)}.
\ee
Finally, we obtain \eqref{elliptic-local-est} by a standard 
iteration procedure working with an increasing sequence of open sets 
$U\cpctsubset V_0\cpctsubset\cdots\cpctsubset V_M\cpctsubset V$
for some $M$ sufficiently large depending only on $s-{n\over 2}$ and
$\sigma-(m-s)$.
\EOProof

The key to proving elliptic estimates on weighted spaces is the following
generalization of Lemma 5.1 of \cite{elliptic-systems-H}.

\lem\label{lem:cc} Let $A$ be a homogeneous constant coefficient linear 
elliptic operator of order $m<n$ on $\Rn$.  For $s\in \Reals$ and 
$\delta\in(m-n,0)$ we have $A :H^s_\delta\rightarrow H^{s-m}_{\delta-m}$
is an isomorphism.
\Proof
We consider three ranges of $s$: $[m,\infty)$, $[-\infty,0]$ and $[0,m]$.

Let $A_{s,\delta}$ denote $A$ acting on $H^s_\delta$.
From \cite{elliptic-systems-H} we know that if $k$ is an
integer and $k\ge m$, then $A_{k,\delta}$ has an inverse $A_{k,\delta}^{-1}$. 
For $s\in[k,k+1]$ we find from interpolation that $A_{k,\delta}^{-1}$ restricts
to a map $B_{s,\delta}:H^{s-m}_{\delta-m}\rightarrow H^s_\delta$, and it 
easily follows that $B_{s,\delta}=A_{s,\delta}^{-1}$.  This establishes 
the result for $s\in[m,\infty)$.

To obtain the result for $s\in(-\infty,0]$ we recall that 
$H^s_\delta=\left(H^{-s}_{-n-\delta}\right)^*$.  Let $A^*$ be
the adjoint of $A$. From the above we know that if $s\le0$
and if $\delta\in(m-n,0)$, then $A^*_{-s+m, -\delta-n+m}$ is an isomorphism.
For $u\in H^{s-m}_{\delta-m}$ let $B_{s,\delta} u$ be the distribution 
defined by
\be
\ip<B_{s,\delta} u,\phi>=\ip<u, (A^*_{-s+m,-\delta+m-n})^{-1}\phi>
\ee
for all $\phi\in C_0^\infty$.  Now
\be
|\ip<B_{s,\delta} u,\phi>| \le ||u||_{H^{s-m}_{\delta-m}} 
||(A^*_{-s+m,-\delta-n+m})^{-1}||_{H^{-s}_{-\delta-n}} 
||\phi||_{H^{-s}_{-\delta-n}}. 
\ee
This proves $B_{s,\delta}u\in H^s_{\delta}$ and we obtain a continuous
map from $H^{s-m}_{\delta-m}\rightarrow H^{s}_{\delta}$.  It easily
follows from the definition of $B_{s,\delta}$ that 
$B_{s,\delta}=A_{s,\delta}^{-1}$.  So the result holds for $s\in(-\infty,0]$.

Finally, the result for $s\in[0,m]$ is obtained by interpolation.
\EOProof

Combining Proposition \ref{prop:elliptic-local} and Lemma \ref{lem:cc} we have
the following mapping property of elliptic operators in $\calL^{m,s}\srho$
on weighted spaces.  The approach of the proof is standard
\cite{global-asympt-simple}
\cite{elliptic-systems-H}\cite{mass-asympt-flat} with some small changes
needed to accommodate the weighted $H^s$ spaces.

\prop\label{prop:extcoercive}  
Suppose $A\in \calL^{m,s}_\rho$ where $s>n/2$, $\sigma\in(m-s,s]$,
and $\rho< 0$.
Then if $m-n<\delta<0$, and $\delta'\in\Reals$ we have
\bel{coercive}
||u||_{H^\sigma_\delta}\lesssim ||A u||_{H^{\sigma-m}_{\delta-m}}
+||u||_{H^{s-m}_{\delta'}}  
\eel
for every $u\in H^\sigma$.
In particular, $A$ is semi-Fredholm as a map from 
$H^\sigma_\delta$ to $H^{\sigma-m}_{\delta-m}$.
\Proof
Let $A=A_\infty+R$ where $A_\infty$ is the principal part of $A$ at
infinity.  Let $\chi$ be a cutoff function such that $1-\chi$ has support
contained in $B_2$ and is equal to 1 on $B_1$.
Let $r$ be a fixed dyadic integer to be selected later,
let $\chi_r(x) = \chi(x/r)$, and let $u_r=\chi_r u$.  From Lemma 
\ref{lem:cc} we have
\be
||u_r||_{H^\sigma\del} \lesssim || A_\infty u_r||_{H^{\sigma-m}_{\delta-m}}.
\ee
Hence
\be
||u_r||_{H^\sigma\del} \lesssim
|| A u_r ||_{H^{\sigma-m}_{\delta-m}} + 
|| R u_r ||_{H^{\sigma-m}_{\delta-m}} 
\ee
where the implicit constant does not depend on $r$.
Now $R\in\calL^{m,s}_\rho$ has vanishing principal part at
infinity.  Hence, from Corollary \ref{cor:opmap} we obtain
\be
|| R u_r ||_{H^{\sigma-m}_{\delta-m}} \lesssim
|| R ||_{H^\sigma_{\delta-\rho}} ||\chi_{r/2}||_{H^s_{-\rho}}
||u_r||_{H^\sigma\del}
\ee
From Lemma \ref{lem:cutoff-scale} proved below we have
\be
\lim_{j\rightarrow\infty} ||\chi_{2^j}||_{H^\sigma_{-\rho}}=0.
\ee
Fixing $r$ large enough we obtain
\bean
||u_r||_{H^\sigma\del} &\lesssim|| A u_r ||_{H^{\sigma-m}_{\delta-m}}
\nonumber\\
&\lesssim 
|| \chi_r A u ||_{H^{\sigma-m}_{\delta-m}} + 
|| [ A,\chi_r] u ||_{H^{\sigma-m}_{\delta-m}} \nonumber\\
&\lesssim 
|| A u ||_{H^{\sigma-m}_{\delta-m}} + 
|| u ||_{H^{\sigma}(B_{2r})}. \label{extest1}
\eean
Let $u_0=(1-\chi_r)u$.  Then 
$|| u_0 ||_{H^\sigma(B_{2r})}\lesssim ||u||_{H^\sigma(B_{2r})}$ and hence
\bea
||u||_{H^\sigma\del} 
&\lesssim || u_r ||_{H^\sigma\del}+|| u_0 ||_{H^\sigma_{B_{2r}}}\\
&\lesssim || A u ||_{H^{\sigma-m}_{\delta-m}} + 
|| u ||_{H^{\sigma}(B_{2r})}
\eea
From Proposition \ref{prop:elliptic-local} we then obtain
\be
||u||_{H^\sigma\del} \lesssim || A u ||_{H^{\sigma-m}_{\delta-m}} + 
|| u ||_{H^{s-m}(B_{3r})}. 
\ee
Equation \eqref{coercive} now follows since for each $\delta'\in \Reals$,
\be
|| u ||_{H^{s-m}(B_{3r})} \lesssim || u ||_{H^{s-m}_{\delta'}}.
\ee
That $A$ is semi-Fredholm is an immediate consequence of \eqref{coercive}
choosing any $\delta'>\delta$.
\EOProof

The following scaling lemma now completes the proof of 
Proposition \ref{prop:extcoercive}.

\lem 
\label{lem:cutoff-scale}
Suppose $f\in H^s_\delta$ with $s\in\Reals$ and $\delta>0$, and suppose
$f$ vanishes in a neighbourhood of the origin.  Then 
\be
\lim_{i\rightarrow\infty} ||T_{-i} f||_{H^s_{\delta}}=0.
\ee
\Proof
Without loss of generality we can assume that $f$ vanishes on $B_2$.
Then $T_{j-i}f=0$ on $B_2$ whenever $j\le i$. So
\bea
||T_{-i}f||_{H^s_{\delta}}^2 &= 
\sum_{j=i+1}^\infty 2^{-2\delta j}||T_j(\phi_j T_{-i} f)||_{H^s}^2\\
&= 2^{-2\delta i} \sum_{j=1}^\infty 2^{-2\delta j}||T^{j}(\phi_j f)||_{H^s}^2\\
&\le 2^{-2\delta i} ||f||_{H^s_{\delta}}^2.
\eea
Since $\delta>0$, the result is proved.
\EOProof

We conclude this section with a result concerning decay properties of
elements in the kernel of elliptic operators in $\calL^{m,s}_\rho$.

\lem\label{lem:kerneldecay}
Suppose $A$ is an elliptic operator in $\calL^{m,s}_\rho$, where 
$s>n/2$ and $\rho< 0$. If
$u\in H^s_{\delta}$ for some $\delta<0$ satisfies $Au=0$, then
$u\in H^s_{\delta'}$ for all $\delta'\in(m-n,0)$.
\Proof
Let $A=A_\infty+R$ where $A_\infty$ is the homogeneous constant coefficient
linear elliptic operator giving the principal part of $A$ at $\infty$.
Then
\be
A_\infty u = - R u \in H^{s-m}_{\delta-m+\rho}.
\ee
Since $A$ is an isomorphism acting on $H^{s}_{\delta'}$ for each 
$\delta'\in(m-n,0)$, we conclude that $u\in H^s_{\delta'}$ for each 
$\delta'\in(\max(m-n,\delta+\rho), 0)$.  Iterating this argument
yields the desired result.
\EOProof

\Remark 
Using partition of unity arguments, the results from 
this section easily extend to asymptotically Euclidean
manifolds. In the sequel, we will use these results 
in the asymptotically Euclidean context without further comment.
\EORemark

\section{The York Decomposition}
\seclabel{sec:york}

In this section we prove that the vector Laplacian is an isomorphism on
certain weighted Sobolev spaces.  The first step is to show it is a 
Fredholm operator with index zero, and that its kernel consists of conformal
Killing fields decaying at infinity.  The second step is to rule out the 
existence of such conformal Killing fields.  These facts are well known
for classical metrics; our only concern is the the low regularity of 
the metrics involved.

To show that the kernel of $\vl$ consists of conformal Killing fields
requires integration by parts, and we need to justify this operation in the
low regularity setting.  Suppose $(M,g)$ is AE of class $H^s_\rho$ with 
$s>n/2$ and  $\rho<0$.  By working in the local charts that define $H^s(M)$, 
it is easy to see that for every $\sigma\in [0,s]$ and 
$\delta\in \Reals$ there is a natural continuous bilinear form
$\ip<\cdot,\cdot>_{(M,g)}: H^{-\sigma}_{\delta}(M)\times H^{\sigma}_{-n-\delta}(M)\ra \Reals$ 
that satisfies
\be
\ip<X,Y>_{(M,g)} = \int_M \ip<X,Y> \dV
\ee
whenever $X$ and $Y$ are smooth compactly supported vector fields.
We note that $\ip<\cdot,\cdot>_{(M,g)}$ can be defined similarly for
scalar functions, and for a fixed pair
$u\in H^{-\sigma}\loc(M)$ and $v\in H^{\sigma}\loc(M)$, the value of
$\ip<u,v>_{(M,g)}$ is also well defined so long as either $u$ or $v$ 
is compactly supported.

Now suppose $X,Y\in H^1\del(M)$.  Since $\vl X \in H^{-1}_{\delta-2}$,
the quantity $\ip<-\vl X,Y>_{(M,g)}$ is well defined if 
$Y\in H^{1}_{2-n-\delta}(M)$ and $s\ge 1$. Since $s>n/2>1$, we only
have the restriction $\delta\le (2-n)/2$.  Under this same condition on 
$\delta$ 
we know $\nabla X, \nabla Y\in L^2(M)$.  For smooth compactly supported 
vector fields
\be
\int_M \ip<\ck X, \ck Y> \dV = \ip<-\vl X, Y>_{(M,g)},
\ee
and a density argument extends integration by parts to 
$X,Y\in H^1_\delta(M)$ when $\delta \le (2-n)/2$.

\prop\label{prop:vlfred}
Suppose $(M,g)$ is AE of class $H^s\srho$ with $s>n/2$ and $\rho<0$.
If $\delta\in(2-n,0)$ then $\vl$ acting on $H^s\del(M)$ is Fredholm 
with index 0.  Moreover, the kernel of $\vl$ is the set of conformal 
Killing fields in $H^s\del$.
\Proof
From Proposition \ref{prop:extcoercive} we know $\vl$ is semi-Fredholm,
and for smooth metrics it is known that $\vl$ has index 0.  Approximating
$g$ with a sequence of smooth metrics $\{ g_k\}$ we have $\vl^k\rightarrow\vl$
and hence the index of $\vl$ is also 0.
Suppose $X\in H^s\del$ and $X\in\ker\vl$.
Then from Lemma \ref{lem:kerneldecay} we know
$X\in H^s_{\delta'}$ for all $\delta'\in(2-n,0)$.  In particular,
we can pick $\delta'\le (2-n)/2$ and integrate by parts to find
\be
0 = \ip<-\vl X, Y>_{(M,g)} =\int_M \ip<\ck X, \ck Y> \dV.
\ee
So $\ck X=0$ in $L^2$ and hence $X$ is a conformal Killing field.  
\EOProof

For smooth metrics on asymptotically Euclidean manifolds we know
\cite{boost-problem} that there are no conformal Killing fields vanishing
at infinity and hence $\vl$ is an isomorphism.  In the case of less
regular metrics, it follows from \cite{bartnik-mfld} Theorem 3.6 that if 
$n=3$ and $g$ is of class $H^s_\rho$ with $s\ge 2$ and $\rho<0$, then there
are no conformal Killing fields in $H^s\del$ for any $\delta<0$.  We
prove now that the same result holds for $s>3/2$.

To do this we consider a boundary value problem.  Let
$\Omega\subset\Rn$ be an open set with smooth boundary and outward
pointing normal $\nu$, and let $\ckflat$ and $\vlflat$ be the
differential operators computed with respect to the Euclidean metric.
The Neumann boundary operator $\Bflat$ for $\vlflat$ takes a vector
field $X$ to the covector field $\ckflat X(\nu,\cdot)$.
From standard elliptic theory we have the following a-priori estimate.
\lem
\label{lem:vlflat-bounded}
Suppose $\Omega$ is a bounded open set with smooth boundary in $\Rn$.
For any vector field $X\in H^s(\Omega)$ with $s>3/2$ we have
\be
||X||_{H^s(\Omega)} \lesssim ||\vlflat X||_{H^{s-2}(\Omega)} 
+ || \Bflat X||_{H^{s-{3\over 2}}(\partial\Omega)} + ||X||_{H^{s-1}(\Omega)},
\ee
where the implicit constant depends on $\Omega$ but not on $X$.
\NoProof
From Lemmas \ref{lem:vlflat-bounded} and \ref{lem:cc} one readily obtains
using the techniques of Proposition \ref{prop:extcoercive} an a-priori 
estimate in the exterior region $E_1$.
\lem\label{lem:vlflat-ext}
For any vector field $X\in H^s\del(E_1)$ with $s>3/2$ and
$\delta\in(2-n,0)$, and for any fixed $\delta'\in\Reals$
\be
||X||_{H^s\del(E_1)} \lesssim ||\vlflat X||_{H^{s-2}_{\delta-2}(E_1)} 
+ || \Bflat X||_{H^{s-{3\over 2}}(\partial E_1)} + ||X||_{H^{s-1}_{\delta'}(E_1)}.
\ee
\NoProof

Using a rescaling argument similar to that in
\cite{maxwell-ahc}, we now show that a conformal Killing field in
$H^s\del$ must vanish identically.  The first step shows that such a
vector field must vanish in a neighbourhood of infinity.

\lem\label{lem:nock-ext} Suppose $(M,g)$ is AE of class $H^s_\rho$ with
$s>n/2$ and $\rho<0$. If $X\in H^s\del$ with $s>n/2$ and
$\delta<0$ is a conformal Killing field, then it vanishes
outside some compact set.
\Proof
We assume for simplicity that $M$ has a single end.
Working in end coordinates, we define a
sequence of metrics $\{g_k\}_{k=1}^{\infty}$ on the exterior region $E_{1}$ 
via $g_k(x) = g(2^kx)$. Since $\rho<0$, it follows from arguments similar
to those of Lemma \ref{lem:cutoff-scale} that $g_k-\gflat$ converges to 0 in 
$H^s_\rho(E_1)$.  It follows that the associated operators 
$\vl^k$, $\ck^k$, and $B^k$ converge to their Euclidean 
analogues as operators on $H^s\del(E_1)$.

Suppose, to produce a contradiction, that $X$ is not identically 0
outside any exterior region $E_R$.  Let $\hat{X}_k(x)=X(2^k x)$ and
let $X_k=\hat{X}_k/||\hat{X}_k||_{H^s\del(E_1)}$.  Since the sequence
$\{X_k\}_{k=0}^\infty$ is bounded in $H^s\del$, we conclude
after reducing to a subsequence that the sequence converges strongly in
$H^{s-1}_{\delta'}$ for any $\delta'\in(\delta,0)$ to some vector field
$X_0$. 

We can assume without loss of generality that $\delta\in (2-n,0)$. 
So from Lemma \ref{lem:vlflat-ext}, the $H^s\del$ boundedness of the sequence 
$\{X_k\}_{k=1}^\infty$, and the identities $\ck^k X_k = 0$, it follows that
\bea
|| X_{k_1}-X_{k_2}||_{H^s\del(E_1)} & \lesssim ||\vlflat
-\vl^{k_1}||_{H^s\del(E_1)} + ||\vlflat -\vl^{k_2}||_{H^s\del(E_1)}
+||\Bflat-B^{k_1}||_{H^s\del(E_1)}\cr &\qquad+
||\Bflat-B^{k_2}||_{H^s\del(E_1)}+||X_{k_1}-X_{k_2}||_{H^{s-1}_{\delta'}(E_1)}.
\eea
We conclude $\{X_k\}_{k=1}^\infty$ is Cauchy in $H^s\del(E_1)$ and
hence converges in $H^s\del$ to $X_0$.  Since $||X_k||_{H^s\del(E_1)}=1$, 
$X_0$ cannot be identically zero. Moreover, since $X_k$ is a conformal 
Killing field
for $g_k$, it follows that $X_0$ is a conformal Killing field for
$\gflat$.  But $\gflat$ does not admit any nontrivial conformal Killing
fields in $H^s\del$, a contradiction.
\EOProof

Using Lemma \ref{lem:nock-ext} together with the a-priori estimate
Lemma \ref{lem:vlflat-bounded}, we now prove that the set of conformal Killing
fields vanishing at infinity is trivial.

\prop\label{prop:nock} Suppose $(M,g)$ is AE of class $H^s_\rho$ with
$s>n/2$ and $\rho<0$. If $X\in H^s_\delta$ with $s>n/2$ and
$\delta<0$ is a conformal Killing field, then it vanishes identically.
\Proof
Let $U$ be the interior of the region where $X$ vanishes.  From
Lemma \ref{lem:nock-ext} we know that $U$ is non-empty.  To show that
$U=M$ it is enough to show that it has empty boundary.  Suppose, to
produce a contradiction, that $x$ is a boundary point of $U$. Working
in local coordinates about $x$, we can assume $M=\Rn$.  Hereafter,
all balls and distances are computed with respect to the flat
background metric. Let $y\in U$ and let $r=d(y,\partial U)$.
Then $B_r(y)\subset U$, and there exists some point $z\in
B_r(y)\cap\partial U$.  After making an affine change of
coordinates, we can assume $z=0$ and $g(0)=\gflat$.

We construct a sequence of metrics $\{g_k\}_{k=1}^\infty$ on
the unit ball by taking $g_k(x) = g(2^{-k}x)$.  
Since $s>n/2$ and $g(0)=\gflat$, it readily follows that
$g_k-\gflat$ converges to 0 in $H^s(B_1)$. It follows that the associated
maps $\vl^k$, $\ck^k$, and $B^k$ converge to their Euclidean counterparts
as operators on $H^s(B_1)$.

We construct vector fields $X_k$ on $B_1$ by setting 
$\hat{X}_k(x)=X(2^{-k}x)$ and letting $X_k=\hat{X}_k/||\hat{X}_k||_{H^s(B_1)}$;
this normalization is possible since $X$ is not identically 0 on $B_{2^{-k}}$.
Since the sequence is bounded in $H^s$, we conclude that the sequence converges
strongly in $H^{s-1}(B_1)$ to some $X_0\in H^s(B_1)$.  Moreover, from the choice
of the point $z$, it follows that $X_k$ vanishes on an open cone
$K$ independent of $k$.  

Arguing as in Lemma \ref{lem:nock-ext}, replacing the use of
Lemma \ref{lem:vlflat-ext} with Lemma \ref{lem:vlflat-bounded}, we conclude
$X_0$ is a conformal Killing field for $\gflat$ and $X_k$ converges in
$H^s(B_1)$ to $X_0$.  In particular, $X_0$ is a nontrivial conformal
Killing field for $\gflat$ on $B_1$ that vanishes on an open cone.
But any such conformal Killing field must vanish identically, a
contradiction.
\EOProof

Combining Propositions \ref{prop:vlfred} and \ref{prop:nock} we 
immediately obtain the York decomposition of symmetric $(0,2)$-tensors.

\thm\label{splitting}  Suppose $(M,g)$ is AE of class $H^s\srho$ with $s>n/2$
and $\rho<0$. Then $\vl$ is an isomorphism acting on $H^s_\delta$ for
any $\delta\in(2-n,0)$. If $S$ is a symmetric trace-free (0,2) tensor
in $H^{s-1}_{\delta-1}$, then there is a unique vector $X\in
H^s_\delta$ and a unique transverse traceless tensor $\sigma\in
H^{s-1}_{\delta-1}$ such that $S = \ck X + \sigma$.
\NoProof

\section{The Linearized Lichnerowicz Equation}
\seclabel{sec:linearlich}

Our existence theorem for solutions of the Lichnerowicz equation
relies on the well-known method of sub and super-solutions.  We
use a variation of the constructive method used by Isenberg
\cite{const-compact} to solve the Lichnerowicz equation on
compact manifolds with $C^2$ metrics. 
The method has subsequently been extended to
weaker classes of metrics \cite{cb-low-reg}\cite{maxwell-ahc}. In this 
section we 
establish properties of the linearized Lichnerowicz equation
$-\Lap+V$ needed to extend the method of sub and super-solutions
to the low regularity setting.

\prop\label{prop:linearprob} Suppose $(M,g)$ is AE of class $H^s\srho$
with  $s>n/2$ and $\rho<0$. Suppose also that $V\in H^{s-2}_{\rho-2}$.  
Then for $\delta\in(2-n,0)$ the operator $-\Lap+V:H^s\del(M)\rightarrow 
H^{s-2}_{\delta-2}(M)$  is Fredholm with index 0.
Moreover, if $V\ge0$ then $-\Lap+V$ is an isomorphism.
\NoProof

We note that if $v\in H^{-\sigma}\loc(M)$ with $\sigma\in [0,s]$, 
we say $v\ge 0$ if
\be
\ip<v,u>_{(M,g)}\ge 0
\ee
for every $u\in H^{\sigma}\loc(M)$ with compact support satisfying
$u\ge0$.  The proof that $-\Lap+V$ has index 0 follows that of Proposition
\ref{prop:vlfred}. The proof of the injectivity of $-\Lap+V$ in the 
case $V\ge 0$ follows from the weak maximum principle proved below.  

\lem\label{lem:maxpr} Suppose $(M,g)$ is AE of class $H^s\srho$ with
$s>n/2$ and $\rho<0$.  Suppose also that $V\in H^{s-2}_{\rho-2}(M)$ and
that $V\ge 0$. If  $u\in H^s\loc$ satisfies
\bel{maxprsys}
-\Delta u +V u  \le 0,
\eel
and if $u^{(+)}=\max(u,0)$ is $o(1)$ on each end of $M$, then $u\le 0$.  
In particular, if $u\in H^s\del(M)$
for some $\delta<0$ and $u$ satisfies \eqref{maxprsys}, then $u\le0$.
\Proof
Fix $\epsilon>0$, and let $v=(u-\epsilon)^{(+)}$.  Since
$u^{(+)}=o(1)$ on each end, we see $v$ is compactly supported.
Moreover, an easy computation shows $uv$ is non-negative, compactly
supported, and belongs to $H^1$.
Since $V\in H^{s-2}_{\rho-2}$, and since $s-2 \ge -1$ we can apply
$V$ to $uv$.  Since $V\ge0 $, and since $uv\ge 0$, we have 
$\ip <V,uv>_{(M,g)} \ge 0$. Finally, since $u$ satisfies \eqref{maxprsys}
we obtain
\be
\ip<-\Delta u,v>_{(M,g)} \le - \ip< V, uv >_{(M,g)} \le 0.
\ee
Integrating by parts we conclude
$v$ is constant and compactly supported, and therefore vanishes identically.
This proves $u\le\epsilon$, and sending $\epsilon$ to 0 proves $u\le 0$.

The statement for $u\in H^s\del(M)$ now follows from the the above and the 
embedding $H^s\del\hookrightarrow C^0\del$.
\EOProof

\lem\label{lem:strmaxpr} Suppose $(M,g)$ is AE of class $H^s\srho$ with
$s>n/2$ and $\rho<0$.  Suppose also that $V\in H^{s-2}_{\delta-2}(M)$
and $u\in H^s\loc(M)$ is nonnegative and satisfies
\be
-\Delta u +V u \ge 0
\ee
If $u(x)=0$ at some point $x\in M$, then $u$ vanishes identically.
\Proof 
Since $M$ is connected, it is enough to show that the
zero set of $u$ is open and closed.  That $u^{-1}(0)$ is closed
is obvious, so we need only verify $u^{-1}(0)$ is open.  We would
like to apply the weak Harnack inequality of \cite{trudinger-measurable}
Theorem 5.2.  
This inequality holds under quite general conditions, and in particular
can be applied to second order elliptic operators of the form
\be
L\;u=\partial_i\left( a^{ij}\partial_j u + a^i u\right) + b^j u_j + a u
\ee
where $a^{ij}$ is continuous, and $a^i\in L^p\loc$, $b^j\in L^p\loc$, and 
$a\in L^{p/2}\loc$ for some $p>n$.  These conditions hold for the Laplacian
when $(M,g)$ is AE of class $H^s$ with $s>n/2$, and also hold for the potential
term $V$ when $n>3$.  When $n=3$, the potential 
term $V$ does not necessarily belong to an $L^p$
space for any $p$, and is not considered directly by 
\cite{trudinger-measurable}.  None-the-less,
a simple transformation of the equation converts it into a more amenable form.

We work in local coordinates near some point $x$ where $u(x)=0$. Let $B$ 
be a small ball 
about $x$ (computed with respect to the flat background metric) and let
$\Phi\in H^s\loc$ be any solution in $B$ of
\bea
\overline{\Lap} \Phi &= V\\
\eea
where $\overline{\Lap}$ is the Laplacian computed with respect to
the flat background metric. It follows that $u$ satisfies
\bel{wharnack}
-\Delta u +\overline{\nabla}\cdot(\overline{\nabla}\Phi u)-
\overline{\nabla}\Phi\cdot\overline{\nabla} u \ge 0
\eel
in a neighbourhood of $x$.  Now
$\overline{\nabla}\Phi\in H^{s-1}\loc$, and
$H^{s-1}\loc\subset L^{q}\loc$ for any $q\ge1$ such that
\be
{1\over q}\ge {1\over 2} - {s-1 \over n}.
\ee
In particular we can pick $q>n$.
Since $u\ge0$ and since $u(x)=0$, the weak Harnack inequality applied
to \eqref{wharnack} implies
$u$ vanishes in a neighbourhood of $x$.
\EOProof

\section{The Cantor-Brill Condition}
\seclabel{sec:lich}

In this section we show that there exists a solution of the Lichnerowicz 
equation
\bel{lichnero}
-a\Lap \phi + R\phi -\abs{\sigma}^2\phi^{-2\dimc-3} = 0
\eel
if and only if the Cantor-Brill condition holds.  We define the
Yamabe invariant
\be
\lambda_g = \inf_{f\in C\cpct^{\infty}(M)\atop f\not\equiv0}
{\int_M a\abs{\nabla f}^2 + R f^2\dV \over ||f||_{L^{2^*}}^2},
\ee
where $2^*=2n/(n-2)$. We note that when $R$ is not a locally integrable 
function, but only a distribution, by $\int_M R f^2\dV$ we mean 
$\ip<R,f^2>_{(M,g)}$.

The proof of the following proposition gives conditions equivalent
to $\lambda_g>0$ and closely follows the analogous result in 
\cite{maxwell-ahc}.

\prop\label{prop:guage} Suppose $(M,g)$ is AE of 
class $H^s\del$ with $s>n/2$, and $\delta\in(2-n,0)$.
Then the following conditions are equivalent:
\bl
\li{1.} There exists a conformal factor $\phi>0$ such that 
$1-\phi\in H^s\del(M)$
and such that $(M,\phi^{4\over n-2}g)$ is scalar flat.
\li{2.} $\lambda_g>0$.
\li{3.} For each $\eta\in[0,1]$, $\calP_\eta=-a\Lap+\eta R$ is an 
isomorphism acting on $H^s\del(M)$.

\el
\Proof
Suppose condition 1 holds.  Since $\lambda_g$ is a conformal invariant,
we can assume that $R=0$.  By solving the equation
\bel{cf1}
-a\Lap v = {\cal R}
\eel
for some smooth positive ${\cal R}\in H^{s-2}_{\delta-2}$,
we can make the conformal change corresponding to $\phi=1+v$ to a metric with 
continuous positive scalar curvature $R$. Let $K$ be the compact core of $M$.
Since
\bel{lps-embed}
||f||_{L^{2^*}}^2\lesssim ||\nabla f||_{L^2}^2 + ||f||_{L^2(K)}^2,
\eel
we find
\be
||f||_{L^{2^*}}^2\lesssim ||\nabla f||_{L^2}^2 + ||R^{1/2} f||_{L^2}^2.
\ee
Hence $\lambda_g>0$.

Now suppose condition 2 holds.  To show condition 3 is true, it is enough 
to show each $\calP_\eta$ has trivial kernel for each $\eta\in[0,1]$.
When $\eta=0$ the result is obvious, so we consider the case $\eta\in(0,1]$.
Suppose, to produce a contradiction, that $\calP_\eta u=0$. 
From Lemma \ref{lem:kerneldecay} we have $u\in H^s_{\delta'}$ for any 
$\delta'\in(2-n,0)$.  Fixing $\delta'\le (2-n)/2$ we can integrate by parts 
to obtain
\bel{yamabe-anal}
0=\int_{M} -au \Lap u + \eta R u^2\dV = 
\int_{M} a\abs{\nabla u}^2 +  \eta R u^2\dV 
\ge \eta \int_{M} a\abs{\nabla u}^2 + R u^2\dV.
\eel
Consider the map $Q(v)= \int_{M} a\abs{\nabla v}^2 + R v^2\dV$ defined
on $H^s_{\delta'}$.  Since $\delta'\le (2-n)/2$, the map $v\mapsto \nabla v$
is continuous from $H^s_{\delta'}$ to $L^2$.  On the other hand, the map
$v\mapsto v^2$ is continuous from $H^s_{\delta'}$ to $H^s_{2\delta'}$.  Since
$R\in H^{s-2}_{\delta-2}$, the map $v->\ip<R,v^2>_{(M,g)}$ is continuous if 
$s> 2 - s $ and $2\delta' \le -n -\delta + 2$.  These last conditions hold
for $s>1$ and $\delta' \le (2-n)/2$.  We have therefore shown that $Q$ is
continuous on $H^s_{\delta'}$.  From \eqref{lps-embed} and the 
inequality $\delta'\le (2-n)/2$ we have the continuous embeddings 
$H^s_{\delta'}\ra H^1_{\delta'}\ra L^{2^*}$.  Thus, approximating $u$ in 
$H^s_{\delta'}$ with smooth compactly supported functions 
$\{u_k\}_{k=1}^\infty$, we find 
\be
{Q(u_k)\over ||u_k||_{L^{2^*}}^2}\ra {Q(u)\over ||u||_{L^{2^*}}^2} \le 0
\ee
by \eqref{yamabe-anal}.  Hence $\lambda_g\le 0$, a contradiction.

Finally, suppose condition 3 holds.  For each $\eta\in[0,1]$, let $v_\eta$
be the unique solution in $H^s_\delta$ of
\be
-a\Lap v_\eta +\eta R v_\eta = -\eta R.
\ee
Setting $\phi_\eta=1+v_\eta$ we see 
\bel{scalarmean2}
-a\Lap \phi_\eta + \eta R \phi_\eta = 0
\eel
To show $\phi_\eta>0$ for each $\eta\in[0,1]$ we use a continuity argument.  
Let $I=\{\eta\in[0,1]:\phi_\eta>0\}$.  The set is open since
$H^s\del \subset C^0\del$, and is nonempty since $\phi_0=1$.
If $\eta_0\in\bar{I}$,  then
$\phi_{\eta_0}\ge 0$.  Since $\phi_\eta$ solves \eqref{scalarmean2},
and since $\phi_{\eta_0}$ tends to 1 at infinity,
Lemma~\ref{lem:strmaxpr} then implies $\phi_{\eta_0}>0$.  Hence $\eta_0\in I$
and $I$ is closed.
  
Letting $\phi=\phi_1$ we have shown $\phi>0$.  
It follows from \eqref{scalarmean2} that $(M,\phi^{4\over n-2}g)$ is 
scalar flat.  Moreover, since $\phi-1\in H^s\del$ it follows from
Lemma \ref{lem:nonlin} and Corollary \ref{cor:nonlin} that
$(M,\phi^{4\over n-2}g)$  is also AE of class $H^s\del$.
\EOProof

In preparation for solving the Lichnerowicz equation, we consider the equation
\bel{fund-eq}
-\Lap u = F(x,u)= {1\over a}\abs{\sigma}^2(1+u)^{-2\kappa-3}.
\eel
A \define{subsolution} of 
\eqref{fund-eq} is a function $u_-$ that satisfies
\bea
-\Lap u_- \le  F(x,u_-),
\eea
and a supersolution $u_+$ is defined similarly with the inequality reversed.

\prop\label{prop:subsup} Suppose $(M,g)$ is AE of class 
$H^s\srho$ with $s>n/2$ and $\rho<0$, and suppose 
$\sigma\in H^{s-1}_{\delta-1}$
with $\delta\in(2-n,0). $ If $u_-,\;u_+\in H^s\del(M)$ are a subsolution and a 
supersolution respectively of \eqref{fund-eq} that satisfy $-1<u_-\le u_+$,  
then there exists a solution 
$u\in H^s\del$ of \eqref{fund-eq} such that $u_-\le u \le u_+$.
\Proof 
\def\LV{\mathop{L_V}}
\def\FV{F_V}
We first treat the case $n=3$ and $3/2 < s \le 2$. For notational 
simplicity, we 
set $c(x) = {1\over a}\abs{\sigma}^2$ and $f(u) = (1+u)^{-2\dimc-3}$
so that $F(x,u)=c(x)f(u)$.
Let $I=[\min u_-,\max u_+]$, and let 
\be
V(x) =c(x)\abs{\min_I f'}.
\ee
It follows that $V\in H^{s-2}_{2\delta-2}$ and is non-negative. 
Let $\FV(x,y)=F(x,y)+V(x)y$, so that $\FV$ is non-decreasing in $y$,
and let $\LV=-\Lap+V$.  Since $V$ is non-negative, $\LV$ 
is an isomorphism acting on $H^s\del$.  

We claim that $||\FV(x,u)||_{H^{s-2}\del}$ is uniformly bounded for 
$u\in H^s\del$ with 
$u_-\le u \le u_+$.  We first consider the term $F(x,u)$.
From Sobolev embedding, $\sigma\in L^q_{\delta-1}$ where
\be
{1\over q} =  {1\over 2} - {s-1\over 3}.
\ee
Hence $\abs{\sigma}^2 \in L^{q/2}_{2\delta-2}$.  If we can show 
$L^{q/2}_{2\delta-2}$ is continuously
embedded in $H^{s-2}_{\delta-2}$, then we will obtain
\bea
||F(x,u)||_{H^{s-2}_{\delta-2}} &\lesssim ||c(x)||_{L^{q/2}_{s-2}} ||f(u)||_{L^\infty} \\
&\lesssim 1.
\eea
Now the dual space of $H^{s-2}_{\delta-2}$ is
$H^{2-s}_{-1-\delta}$.  From Sobolev embedding, 
$H^{2-s}_{-1-\delta}$ is continuously embedded in $L^p_{-1-\delta}$, where
\be
{1\over p} = {1\over 2} - {2-s\over 3}.
\ee
The dual space of $L^p_{-1-\delta}$ is $L^{p'}_{\delta-2}$ where
\be
{1\over p'} = {1\over 2} + {2-s\over 3}.
\ee
Since $q/2\ge p'$ exactly when $s\ge 3/2$, and since $2\delta-2 < \delta-2$,
we obtain the continuous embedding of $L^{q/2}_{2\delta-2}\hookrightarrow H^{s-2}_{\delta-2}$
as desired. A similar analysis shows that 
$||V(x)u||_{H^{s-2}_{\delta-2}} \lesssim 1$ and we obtain
\bel{subsupe1}
||\FV(x,u)||_{H^{s-2}_{\delta-2}}\lesssim 1.
\eel

We now construct a monotone decreasing sequence of functions 
$u_+=u_0\ge u_1 \ge u_2 \ge \cdots$ by iteratively solving
\be
\LV u_{i+1} = \FV(x,u_i).
\ee
The monotonicity of the sequence follows from the maximum principle 
Lemma \ref{lem:maxpr}
and the monotonicity of $\FV(x,y)$ in $y$.  The maximum
principle also implies $u_i\ge u_-$.

From Proposition \ref{prop:linearprob} we can estimate
\bel{subsupe2}
||u_{i+1}||_{H^s\del}
\lesssim ||\FV(x,u_{i})||_{H^{s-2}_{\delta-2}}. 
\eel
From \eqref{subsupe1} and \eqref{subsupe2} we 
conclude the sequence $\{u_i\}_{i=1}^\infty$ is bounded in $H^s\del$.
In particular, a subsequence of $\{u_i\}_{i=1}^\infty$ (and by monotonicity, the 
whole sequence) converges weakly in $H^s\del$ to a limit $u_\infty$. 

It remains to see
$u_\infty$ is a solution of \eqref{fund-eq}.
Now $u_i$ converges strongly to $u_\infty$ in $H^{s'}_{\delta'}$ for any 
$s'<s$ and $\delta'>\delta$.  Hence, on any compact set the sequence
converges in $H^1$ and in $C^0$. We conclude that for any fixed
$\phi\in C\cpct^\infty$, each $f_j(u_i)\phi$ converges in $H^1$ to 
$f_j(u_\infty)\phi$.  Since each $c_j\in H^{s-2}_{\rho-2}$, and
since $s-2>-1$, we find
\be
\int_M\left(\FV(x,u_i)-V(x)u_{i+1}\right)\phi\dV \rightarrow 
\int_M F(x,u_\infty)\phi\dV.
\ee
Moreover, 
\be
\int_M \ip<\nabla u_{i+1}, \nabla \phi> \dV
\rightarrow
\int_M \ip<\nabla u_\infty, \nabla \phi>\dV.
\ee
Hence
\be
\int_M \ip<\nabla u_\infty, \nabla \phi>\dV =
\int_M F(x,u_\infty)\phi\dV. 
\ee
Since $\phi\in C\cpct^\infty$ is arbitrary, we conclude 
$-\Lap u_\infty + F(x,u_\infty)=0$ as a distribution.

To handle the case $n=3$ and $s > 2$ we use a bootstrap.  First
suppose $4\ge s \ge 2$.  From the above we have a solution $u$ in 
$H^2_\delta$. Since $2 > n/2=3/2$ and since $2>s-2 \in [0,2]$, we know
from Lemma \ref{lem:nonlin} and the remark following it that 
$c(x)f(u)\in H^{s-2}_{\delta-2}$.  Since $-\Lap u \in H^{s-2}_{\delta-2}$, 
Proposition \ref{prop:linearprob} implies $u\in H^s_\delta$. We
obtain the result for all $s>3/2$ by induction.

Finally, we turn to the case $n>3$.  Here we cannot start a bootstrap from
$s=2$ since $H^2$ is not an algebra.  Instead we rely on results from
\cite{maxwell-ahc} for metrics of class $\W kp\del$.  Let $k$ be the greatest 
integer such that $k\le s$.  From Sobolev embedding we find $(M,g)$ is
asymptotically Euclidean of class $\W kp_\rho$ where 
\be
{1\over p} = {1\over 2} - {s-k \over n}.
\ee
Since $n>3$ it easily follows that $2 \le p < \infty$ and moreover that
$k\ge2$ and $k>n/p$.  Similarly, 
$c(x)\in \W {k-2}p_{\delta-2}$.  From \cite{maxwell-ahc}
we conclude that there exists a solution $u\in \W kp\del$.  An easy computation
shows multiplication by $f(u)$ is a continuous involution of $H^{k-1}_{\delta-2}$ 
and $H^{k-2}_{\delta-2}$. From interpolation we conclude 
$c(x)f(u)\in H^{s-2}_{\delta-2}$.
Proposition \ref{prop:linearprob} then implies there exists 
$\hat u\in H^s_\delta$ with $-\Lap \hat u = c(x)f(u)$.  But 
$\hat u\in \W kp\del$ and the Laplacian of a $\W kp_\rho$ metric
is an isomorphism of $\W kp\del$.  Hence $\hat u = u$ and we obtain a solution
in $H^s\del$.
\EOProof

Using Proposition \ref{prop:subsup} we readily obtain the existence of solutions
of \eqref{lichnero}.

\thm \label{thm:main} Suppose $(M,g)$ is AE of 
class $H^s\del$ with $s>n/2$ and $\delta\in(2-n,0)$.
Let $\sigma$ be any transverse-traceless tensor in $H^{s-1}_{\delta-1}(M)$. 
There exists a conformal factor $\phi$ solving \eqref{lichnero}
if and only if $\lambda_g>0$.  Moreover, if a solution exists then it
is unique.
\Proof
If a solution exists, then it follows from \eqref{lichnero} that
$g$ is conformally related to a metric with non-negative scalar curvature,
and from Proposition \ref{prop:guage} that $\lambda_g>0$.

If $\lambda_g>0$ we can assume without loss of generality that 
$R=0$.  Setting $\phi = 1+v$, solving the Lichnerowicz equation 
is equivalent to solving
\bel{lichreduced}
-a\Lap v = \abs{\sigma}^2(1+v)^{-2\dimc-3}
\eel
with the constraint $v>-1$.  

Evidently $v_-=0$ is a subsolution  of \eqref{lichreduced}. To find a 
supersolution, we solve 
\bel{supeq}
-a\Lap v_+= \abs{\sigma}^2.
\eel
From the weak maximum principle we find $v_+\ge v_- = 0 $.  We can now 
apply Proposition \ref{prop:subsup}
to find there exists a nonnegative solution $v$ of \eqref{lichreduced} in 
$H^s \del$.  

The uniqueness of the solution follows from the fact that $(1+v)^{-2\dimc-3}$ 
is decreasing in $v$ for $v>-1$ and a maximum principle argument
similar to that of Lemma \ref{lem:maxpr}.
\EOProof

\section{Approximation by Smooth Solutions}
\seclabel{sec:approx}

The following theorem shows that a rough, maximal, AE solution of the 
constraints can be approximated arbitrarily well by smooth solutions.

\thm\label{thm:approx} Let $(M,g_0,K_0)$ be 
a maximal AE solution
of the constraint equations of class $H^s_{\delta}$ with $s>n/2$ and
$\delta\in(n-2,0)$. For
any $\epsilon>0$, there exists a maximal AE 
solution $(M,g_\epsilon,K_\epsilon)$ of the constraint
equations of class $H^t\del$ for every $t\ge s$
such that 
$||g_0-g_\epsilon||_{H^s_\delta}<\epsilon$ and 
$||K_0-K_\epsilon||_{H^{s-1}_{\delta-1}}<\epsilon$.
\Proof Let $\{g_k\}_{k=1}^\infty$ be a sequence of metrics on $M$ 
in $H^t_\delta$ for every $t\ge s$ such that 
$||g_k-g_0||_{\Hmt}\ra 0$.  We will write $\vl^k$, $\ck^k$, and $\divk$
for the differential operators corresponding to $g_k$.

To construct a sequence of transverse-traceless tensors, we
let $\{S_k\}_{k=1}^\infty$ be an arbitrary  sequence of 
traceless $(0,2)$-tensors in $H^{t-1}_{\delta-1}$ for every $t\ge s$
converging to $K_0$ in $\Htt$.  
Let $X_k\in \Hmt$ be the unique solution of $\vl^k X_k=\divk S_k$.  
Since $\divk S_k\in H^{t-2}_{\delta-2}$ for every $t\ge s$,
it follows that $\ck X_k\in H^{t-1}_{\delta-1}$ for every $t\ge s$.
Since $\vl$ is invertible, we have uniform bounds on the norm of the 
inverse of $\vl^k$. Hence
\be
||X_k||_{\Hmt}\lesssim ||\divk S_k|| \lesssim 
||\divk||_\Htt ||S_k- K_0||_\Htt + 
||\divk-\div||_\Htt|| K_0 ||_\Htt.
\ee
So $||X_k||_\Hmt\ra 0$.  Letting $\sigma_k= S_k-\ck^k X_k$
it follows that $\sigma_k$ is transverse-traceless with respect
to $g_k$, and is in $H^{t-1}_{\delta-1}$ for every $t\ge s$.  
Moreover, $||\sigma_k- K_0||_\Htt \le 
||\ck^k X_k||_\Htt + ||S_k- K_0||_\Htt \ra 0$.

The correction to $g_k$ is now accomplished by the implicit function 
theorem. Let
\be
\calF(g,\sigma,v) = -a\Delta_g v+R_g(1+v)-\abs{\sigma}^2_g(1+v)^{-2\dimc-3}.
\ee
For fixed $g$ and $\sigma$, let $\calF_{g,\sigma}(v)= \calF(g,\sigma,v)$.
Using Lemma \ref{lem:deriv} proved below, we find that
the Fr\'echet derivative of $\calF_{g,\sigma}$ is
\be
\diff \calF_{g,\sigma}(v)(h) =
-a\Delta_g h +R_g h +(2\dimc+3)\abs{\sigma}^2_g(1+v)^{-2\dimc-4}h,
\ee
and $\diff\calF_{g,\sigma}(v)$ is hence continuous
in a neighbourhood of $(g,\sigma,v)$ for each $v$ with $v>-1$.  Moreover,
\be
\diff\calF_{g_0,K_0}(0)(h) = L(h)=-a\Delta_{g_0} h+R_{g_0} h+
(2\dimc+3)\abs{K_0}^2 h.
\ee
Since $R_0$ is nonnegative, $L$ is an isomorphism.  
Since $\calF(g_0,K_0,0)=0$, and since $g_k\ra g_0$ and $\sigma_k\ra K_0$,
the implicit function theorem (e.g. \cite{primer-nonlin-analysis} Lemma 2.2.1)
implies that for $k$ sufficiently large
there exists $v_k\in\Hmt$ such that $v_k\ra 0$ and such that 
$\calF(g_k,\sigma_k,v_k)=0$. From the equation $\calF(g_k,\sigma_k,v_k)=0$
and a bootstrap we have $v_k\in H^t_\delta$ for every $t\ge s$.
Letting $\hat g_k = (1+v_k)^{2\dimc} g_k$ and $\hat K_k=(1+v_k)^{-2}\sigma_k$, 
we conclude from Lemma \ref{lem:nonlin} and 
Corollary \ref{cor:nonlin} that $(M,\hat g_k,\hat K_k)$
is an AE data set of class $H^t_\delta$ for every $t\ge s$ and
$(\hat g_k-g_0,\hat K_k-K_0)$ converges to $0$ in $\Hmt\times\Htt$.
Taking $k$ sufficiently large proves the theorem.
\EOProof

The following lemma completes the proof of Theorem \ref{thm:approx}.
\lem\label{lem:deriv}
Suppose $(M,g)$ is AE of class $H^s_\rho$ with $s>n/2$ and 
$\rho <0$, and suppose $\sigma\in H^{s-1}_{\delta-1}$ with 
$\delta\in(2-n,0)$.  Let $U$ be the open subset $\{v\in H^s_\delta: v>-1\}$,
and let 
$\calG: U\ra H^{s-2}_{\delta-2}$ be given by
\be
\calG(v) = \abs{\sigma}^2 (1+v)^{-2\dimc-3}.
\ee
Then $\calG$ has a Fr\'echet derivative $\diff \calG$ given by
\be
\diff \calG(v)(h) =  -(2\dimc+3) \abs{\sigma}^2 (1+v)^{-2\dimc-4}h.
\ee
\Proof
We first consider the maps
\bea
 g(v) &= (1+v)^{-2\dimc-3}\\
g'(v) &= -(2\dimc+3)(1+v)^{-2\dimc-4}\\
\eea
Since $1\in H^s_\epsilon$ for every $\epsilon>0$, it follows from Lemma
\ref{lem:nonlin} that $g$ and $g'$ are continuous as maps from $U$ to
$H^s_\epsilon$.  Since $1+v>0$, it follows that there exists a ball $B_r$ 
of radius $r$ in $H^s_\delta$ such that $1+v+h>0$ for all $h\in B_r$.  Taking
$h\in B_r$, it follows from the continuity of $g'$ that the map 
$t\ra g'(v+th)h$ from $[0,1]$ to $H^s_\epsilon$ is Bochner integrable. It
is easy to see that
\be
g(v+ h)-g(v) = \int_0^1 g'(v+ t h)h\,dt,
\ee
where the integral is a Bochner integral.  Now
\bea
|| g(v+ h)-g(v) - g'(v)h ||_{H^s_\epsilon} &=
\bnorm{ \int_0^1 \left(g'(v+t h)-g'(v)\right) h \,dt }_{H^s_\epsilon}\\
&\le
\int_0^1 \bnorm{ \left(g'(v+ th)-g'(v)\right) h }_{H^s_\epsilon}dt\\
&\le \int_0^1 \bnorm{ \left(g'(v+th)-g'(v)\right)}_{H^s_\epsilon} dt\, ||h||_{H^s_\delta}.
\eea
Since $\abs\sigma^2\in H^{s-2}_{2\delta-2}$, we can take $\epsilon < -\delta$
to obtain
\be
|| \calG(v+th)-\calG(v) - \diff\calG(v)h ||_{H^{s-2}_{\delta-2}}\lesssim
||\sigma^2||_{H^{s-2}_{2\delta-2}} 
\int_0^1 \bnorm{ \left(g'(v+th)-g'(v)\right)}_{H^s_\epsilon} dt\, 
||h||_{H^s_\delta}.
\ee
Since $g'$ is continuous in a neighbourhood of $v$, it follows that
\be
\int_0^1 \bnorm{ \left(g'(v+th)-g'(v)\right)}_{H^s_\epsilon} dt 
\ee
can be made arbitrarily small by taking $||h||_{H^s_\delta}$ small.
We conclude that $\diff\calG(v)$ is the Fr\'echet derivative of $\calG$ at $v$.
\EOProof

\Acknowledge
I would like to thank D. Pollack and H. Smith for helpful conversations. The 
research for this paper was partially supported by NSF grant DMS-0305048.
\EOAcknowledge

\references
\ifjournal
\bibliographystyle{journal}%
\else
\advance\baselineskip by -2 pt
\bibliographystyle{amsalpha-abbrv}%
\fi
\bibliography{mrabbrev,ahconstraints,gep,rough}
\endmatter